%% file: main.tex
\documentclass[11pt,letterpaper]{article}
\input{preamble}

\title{Constant Approximation of Fr\'echet Distance in Near-Linear Time}
\input{authors}
\date{}

\begin{document}

\pagenumbering{roman}
\maketitle
\input{abstract}
\thispagestyle{empty}
\clearpage

\tableofcontents
\clearpage

\pagenumbering{arabic}
\input{sections/01-introduction}

\input{sections/02-preliminaries}
\input{sections/03-technique-overview}
\input{sections/04-boundary-maps}
\input{sections/05-query-recursion}
\input{sections/06-index-construction}
\input{sections/07-parameters}
\input{sections/08-conclusion}
\input{sections/ai-methodology}

\appendix
\input{sections/A-threshold-search}
\input{sections/B-discrete}
\input{sections/C-related-work}

\phantomsection
\addcontentsline{toc}{section}{References}
\bibliographystyle{alphaurl}
\bibliography{references}

\end{document}

%% file: preamble.tex
\usepackage[utf8]{inputenc}
\usepackage[T1]{fontenc}
\usepackage{lmodern}
\usepackage[letterpaper,margin=1in]{geometry}
\usepackage{amsmath,amssymb,amsthm,mathtools}
\usepackage{microtype}
\usepackage{enumitem}
\usepackage{booktabs,tabularx}
\usepackage{algorithm}
\usepackage{algpseudocode}
\usepackage[section]{placeins}
\usepackage{tikz}
\usetikzlibrary{positioning}
\usepackage{titlesec}
\usepackage[titles]{tocloft}
\usepackage[hidelinks,unicode,pdfusetitle]{hyperref}

\titlespacing*{\section}{0pt}{2.6ex plus .8ex minus .3ex}{1.4ex plus .2ex}
\titlespacing*{\subsection}{0pt}{2.1ex plus .6ex minus .2ex}{1.1ex plus .2ex}
\makeatletter
\g@addto@macro\normalsize{%
  \setlength\abovedisplayskip{6pt plus 2pt minus 3pt}%
  \setlength\belowdisplayskip{6pt plus 2pt minus 3pt}%
  \setlength\abovedisplayshortskip{2pt plus 2pt}%
  \setlength\belowdisplayshortskip{4pt plus 2pt minus 2pt}}
\makeatother
\definecolor{figblue}{HTML}{0072B2}
\definecolor{figred}{HTML}{C43C39}
\definecolor{figorange}{HTML}{D58B00}
\definecolor{figgreen}{HTML}{00966C}

\newcommand{\eps}{\varepsilon}
\newcommand{\R}{\mathbb R}
\newcommand{\tO}{\widetilde O}
\newcommand{\dF}{d_{\mathrm F}}
\newcommand{\ddF}{\tilde{d}_{\mathrm{F}}}
\newcommand{\Build}{\mathsf{Build}}
\newcommand{\SelfQuery}{\mathsf{SelfQuery}}
\newcommand{\Query}{\mathsf{Query}}

\newcommand{\Decide}{\mathsf{Decide}}
\newcommand{\In}{\mathrm{In}}
\newcommand{\Out}{\mathrm{Out}}
\newcommand{\Lift}{\operatorname{Lift}}
\newcommand{\rank}{\operatorname{rank}}
\newcommand{\Rr}[1]{\mathcal R_{#1}}
\algrenewcommand\algorithmicrequire{\textbf{Input:}}
\algrenewcommand\algorithmicensure{\textbf{Output:}}

\newtheoremstyle{plaintight}{5pt plus 1pt minus 1pt}{5pt plus 1pt minus 1pt}%
  {\itshape}{}{\bfseries}{.}{.5em}{}
\newtheoremstyle{deftight}{5pt plus 1pt minus 1pt}{5pt plus 1pt minus 1pt}%
  {\normalfont}{}{\bfseries}{.}{.5em}{}
\numberwithin{equation}{section}
\theoremstyle{plaintight}
\newtheorem{theorem}{Theorem}[section]
\newtheorem{lemma}[theorem]{Lemma}

\newtheorem{corollary}[theorem]{Corollary}
\theoremstyle{deftight}
\newtheorem{definition}[theorem]{Definition}
\theoremstyle{remark}

\setlist{itemsep=2pt,topsep=4pt,parsep=0pt}

%% file: authors.tex
\author{%
  Haoqiang Huang\thanks{Huawei, Beijing, China. 
    Email: \texttt{huanghaoqiang@huawei.com}.}
 \and
 Zihe Wang\thanks{Renmin University of China, Beijing, China.
   Email: \texttt{wang.zihe@ruc.edu.cn}.}
 \and
 Shuo Zhang\thanks{Renmin University of China, Beijing, China.
   Email: \texttt{zhangshuo1422@ruc.edu.cn}.}
}

%% file: abstract.tex

\begin{abstract}
\normalsize
We give the first constant-factor approximation for the Fr\'echet
distance in near-linear time.
Given two polygonal curves in $\R^d$ with $n$ and $m$ vertices,
respectively, where $m\le n$, our deterministic algorithm runs in
$O(n^{1+\eps})$ time for every fixed $\eps>0$ and $d$, with an
approximation factor depending only on $\eps$ and $d$. The same result holds for
the discrete Fr\'echet distance.
Our result extends the line of work initiated in~\cite{CHZ} from
strongly subquadratic to near-linear time.

Our algorithm recursively balances preprocessing and query time.
An index for comparing subcurves of a single curve supports faster
queries between different curves. We obtain this speedup by replacing
one curve with surrogates assembled from indexed subcurves of the other
and shared across starting points of matchings. Faster queries then
reduce the cost of building the next index. Compact boundary
reachability summaries allow the answers to be reused in later
recursive calls. After a number of rounds depending only on $\eps$,
both preprocessing and query time are near-linear.
\end{abstract}

%% file: sections/01-introduction.tex

\section{Introduction}
\label{sec:intro}

\emph{Fr\'echet distance} is a natural and widely used measure of
similarity between curves~\cite{Alt09}. It has applications in
spatio-temporal data mining, including trajectory clustering and map
construction~\cite{BBGL08,BBG20}.

Let $P=(p_0,\ldots,p_{n-1})$ and $Q=(q_0,\ldots,q_{m-1})$ be
polygonal curves in $\R^d$, where $m\le n$ and $d$ is fixed. A
\emph{parameterization} of $P$ is a continuous function
$\rho:[0,1]\to P$ such that $\rho(0)=p_0$, $\rho(1)=p_{n-1}$,
and $\rho(t)$ does not occur after $\rho(t')$ along $P$ whenever
$t\le t'$. Let $\sigma$ be a parameterization of $Q$. The pair
$M=(\rho,\sigma)$ defines a \emph{matching} of $P$ and $Q$, with
$\rho(t)$ matched to $\sigma(t)$ for every $t\in[0,1]$. Its
\emph{cost} and the \emph{Fr\'echet distance} are
\[
 \operatorname{cost}(M)=\max_{t\in[0,1]}\|\rho(t)-\sigma(t)\|,
 \qquad
 \dF(P,Q)=\inf_M\operatorname{cost}(M),
\]
where $\|\cdot\|$ is the Euclidean norm and the infimum ranges over
all matchings. For polygonal curves, the infimum is attained. The
\emph{discrete Fr\'echet distance} $\ddF(P,Q)$ uses sequences of
vertex pairs from $(p_0,q_0)$ to $(p_{n-1},q_{m-1})$, advancing to
the next vertex on one or both curves at each step. It is the minimum,
over such sequences, of the largest distance between paired vertices.

The classical algorithms take roughly quadratic time:
Alt and Godau~\cite{AG95} compute $\dF(P,Q)$ in $O(nm\log(nm))$
time using a geometric reachability formulation, and Eiter and
Mannila~\cite{EM94} compute $\ddF(P,Q)$ in $O(nm)$ time by dynamic
programming. Subsequent work has obtained logarithmic improvements in
exact computation and the corresponding decision
problems~\cite{AAKS13,BBMM14,CH25,CH26}. A different improvement is
possible for the Fr\'echet distance in one dimension: Blank and
Driemel~\cite{BD24} give $\tO(n+m^2)$ time, which is particularly
effective when one curve is much shorter than the other.

Conditional lower bounds help explain the difficulty of obtaining
larger improvements for general curves. Bringmann~\cite{Bri14} showed
that, assuming the Strong Exponential Time Hypothesis
(SETH)~\cite{IP01},\footnote{In October 2026, a claimed refutation
of the Exponential Time Hypothesis was released for independent
verification~\cite{OpenAI26}. If confirmed, it would also refute SETH.
The lower bounds discussed here are obtained by reductions from the
Orthogonal Vectors problem~\cite{Bri14,BM16,BOS19,Blank26}, so they
also hold under the Orthogonal Vectors Hypothesis (OVH). SETH implies
OVH~\cite{Wil05}, but a refutation of SETH alone would not refute OVH.}
neither version admits an $O((nm)^{1-\gamma})$-time algorithm for any
constant $\gamma>0$ in dimension at least two. For two $n$-vertex
curves, this rules out \emph{strongly subquadratic} time, that is,
$O(n^{2-\gamma})$ for a constant $\gamma>0$. Bringmann and
Mulzer~\cite{BM16} established such a lower bound for the discrete
distance even in one dimension. Buchin, Ophelders, and
Speckmann~\cite{BOS19} further ruled out strongly subquadratic
approximation within any fixed factor below $3$, for both versions
already on one-dimensional curves of equal size. Blank~\cite{Blank26}
extended approximation lower bounds to curves of different sizes,
with the excluded factors depending on the dimension and metric.

Approximation nevertheless permits much faster algorithms when the
curves have additional geometric structure. Driemel, Har-Peled, and
Wenk~\cite{DHW12} introduced $c$-packed curves, whose length inside any
ball of radius $r$ is at most $cr$, and gave a $(1+\eta)$-approximation
in near-linear time for fixed $c$ and $\eta>0$. Bringmann and
K\"unnemann~\cite{BK17} improved the dependence on $\eta$, and linear
time for fixed $c$ and $\eta$ was recently
achieved~\cite{BDHR26}. For arbitrary curves, obtaining even a
constant-factor approximation proved more difficult. A series of
$O(\alpha)$-approximations traded accuracy for
speed~\cite{BM16,CR18,CF21,HKS23,HO24,HKS25}, but for two $n$-vertex
curves, polynomial savings in running time required approximation
factors polynomial in $n$.

Cheng, Huang, and Zhang~\cite{CHZ} gave the first
constant-factor approximation in strongly subquadratic time, namely a
randomized $(7+\eta)$-approximation for both versions that runs in
$O(nm^{0.99}\log n)$ time and succeeds with high probability.
Here and below, $\eta>0$ is fixed independently of the input size.
Liu and Wang~\cite{LW26} improved the factor to $5+\eta$, with
running times $\tO(nm^{8/9})$ for the Fr\'echet distance and
$\tO(nm^{4/5})$ for the discrete Fr\'echet distance. Van der Horst and
Ophelders~\cite{HO} obtained a deterministic $(3+\eta)$-approximation
in $\tO(nm^{2/3})$ time for both versions, nearly matching the
conditional lower bound on the approximation factor. For curves of
very different sizes, Blank~\cite{Blank26} gave a complementary
$(3+\eta)$-approximation in $\tO(n+m^2)$ time, which is near-linear
when $m\le\sqrt n$. Table~\ref{tab:comparison} summarizes exact and
approximation bounds, and Appendix~\ref{sec:related-work} gives a
fuller discussion of the related work and its connections to our
techniques.

\begin{table}[!htb]
\centering
\footnotesize
\setlength{\tabcolsep}{4pt}
\renewcommand{\arraystretch}{1.13}
\begin{tabularx}{\textwidth}{>{\centering\arraybackslash}p{2.0cm}|>{\centering\arraybackslash}p{3.4cm}|>{\centering\arraybackslash}X|>{\centering\arraybackslash}p{2.6cm}}
Setting & Solution & Running time & Reference\\
\hline
$\dF$ & Exact & $O((n^2m+nm^2)\log(nm))$ & \cite{God91}\\
$\ddF$ & Exact & $O(nm)$ & \cite{EM94}\\
$\dF$ & Exact & $O(nm\log(nm))$ & \cite{AG95}\\
$\ddF$, $d=2$ & Exact (word RAM) & $O(nm\log\log n/\log n)$ & \cite{AAKS13}\\
$\dF$, $d=2$ & Exact (pointer machine) & $O(nm\sqrt{\log n}(\log\log n)^{3/2})$ expected & \cite{BBMM14}\\
$\dF$, $d=2$ & Exact (word RAM) & $O(nm(\log\log n)^2)$ expected & \cite{BBMM14}\\
$\dF$, $d=1$ & Exact & $O(m^2\log^2 n+n\log n)$ & \cite{BD24}\\
$\dF$ & Exact ($0<\mu<1$) & $O\!\left(\frac{nm(\log\log n)^{2+\mu}\log n}{\log^{1+\mu}m}\right)$ expected & \cite{CH25}\\
$\ddF$, $m=n$ & Exact & $O((n^2/\log^2 n)\log^2\log n)$ & \cite{CH26}\\
$\dF$, $m=n$ & Exact & $O((n^2/\log^2 n)\log^{O(1)}\log n)$ expected & \cite{CH26}\\
\hline
$\ddF$, $m=n$ & $O(\alpha)$-approx. & $O(n\log n+n^2/\alpha)$ & \cite{BM16}\\
$\ddF$, $m=n$ & $O(\alpha)$-approx. & $O(n\log n+n^2/\alpha^2)$ & \cite{CR18}\\
$\dF$, $m=n$ & $O(\alpha)$-approx. & $O((n^3/\alpha^2)\log n)$ & \cite{CF21}\\
$\dF$ & $O(\alpha)$-approx. & $O((n+nm/\alpha)\log^3 n)$ & \cite{HKS23}\\
$\dF$, $m=n$ & $O(\alpha)$-approx. & $O((n^2/\alpha)\log^2 n)$ & \cite{HO24}\\
$\dF$, $m=n$ & $O(\alpha)$-approx. & $O((n^2/\alpha)\log n)$ & \cite{HKS25}\\
\hline
$\dF$, $\ddF$ & $(7+\eta)$-approx. & $O(nm^{0.99}\log n)$ with high probability & \cite{CHZ}\\
$\dF$ & $(5+\eta)$-approx. & $\tO(nm^{8/9})$ with high probability & \cite{LW26}\\
$\ddF$ & $(5+\eta)$-approx. & $\tO(nm^{4/5})$ with high probability & \cite{LW26}\\
$\dF$, $\ddF$ & $(3+\eta)$-approx. & $\tO(n+m^2)$ & \cite{Blank26}\\
$\dF$, $\ddF$ & $(3+\eta)$-approx. & $\tO(nm^{2/3})$ & \cite{HO}\\
\hline
$\dF$, $\ddF$ & $f(1/\eps)$-approx. & $O(n^{1+\eps})$ & Theorem~\ref{thm:main}
\end{tabularx}
\caption{Previous results.}
\label{tab:comparison}
\end{table}

Despite this series of previous results, no constant-factor
approximation in near-linear time was known for arbitrary input
sizes. This raises the following question.
\begin{quote}
\emph{Is there a constant-factor approximation algorithm for the
Fr\'echet distance that runs in near-linear time?}
\end{quote}
For edit distance, the analogous question has a positive answer.
Chakraborty, Das, Goldenberg, Kouck\'y, and
Saks~\cite{CDGKS18} gave the first constant-factor approximation of
edit distance in truly subquadratic time, and Andoni and Shekel
Nosatzki~\cite{AN} subsequently obtained a constant-factor
approximation in $O(n^{1+\eps})$ time for every fixed $\eps>0$.
    
\paragraph{Our results.}
We show that the answer to the question above is positive, for both
the Fr\'echet distance and the discrete Fr\'echet distance. For every fixed
$\eps>0$, our algorithm runs in $O(n^{1+\eps})$ time and
approximates the distance within a factor that depends only on $\eps$
and on the dimension. 

\begin{theorem}[Main result]
\label{thm:main}
For every fixed dimension $d\ge1$, there is a function
$f:(0,\infty)\to[1,\infty)$ such that, for every $\eps>0$, a
deterministic algorithm, given two nonempty polygonal curves $P,Q$
in $\R^d$ with $n,m$ vertices, where $m\le n$, computes in
$O(n^{1+\eps})$ time a value
$\widehat D$ such that $\dF(P,Q)\le\widehat D\le f(1/\eps)\cdot\dF(P,Q)$.
The same guarantee holds for the discrete Fr\'echet distance $\ddF(P,Q)$.
\end{theorem}

Throughout the paper,
\emph{near-linear} refers to this $O(n^{1+\eps})$ guarantee, with
$\eps$ fixed independently of the input. We show that
$f(1/\eps)\le\exp(\exp(O(\eps^{-1}\log(2/\eps))))$ for $0<\eps\le1$
(Lemma~\ref{ix:parameters}). A doubly exponential dependence on
$1/\eps$ also appears in the near-linear-time approximation for
edit distance of Andoni and Shekel Nosatzki~\cite{AN}.

%% file: sections/02-preliminaries.tex

\section{Preliminaries}
\label{sec:preliminaries}

Let $P$ be a polygonal curve in $\R^d$ with vertices
$p_0,\ldots,p_{n-1}$, and let $|P|=n$ be its size. We parameterize $P$
over $[0,n-1]$, so that $P(i)=p_i$ and $P$ is affine on each interval
$[i,i+1]$. Consecutive vertices may coincide. Take two points $a, a' \in P$, we say $a\le_P a'$
if  $a$ is not behind $a'$ along $P$. For
$a,a'\in[0,n-1]$ with $a\le_P a'$, we use $P[a,a']$ to denote the subcurve between $a$ and $a'$.
Throughout, $Q$ is a
second polygonal curve with $m$ vertices, parameterized over $[0,m-1]$,
the dimension $d$ is a fixed constant, and $\|\cdot\|$ is the Euclidean
norm.

We first develop approximate decision procedures. As in~\cite{CF21,BM16},
a search over the radius then yields approximation algorithms for the
Fr\'echet and discrete Fr\'echet distances with only a polylogarithmic
overhead in running time (Lemma~\ref{ix:search}). For $r\ge0$, a matching of curves $X,Y$ with cost at most $r$ is an
\emph{$r$-matching}, or a \emph{witness} for $\dF(X,Y)\le r$.
For $C\ge1$, a \emph{$C$-approximate decision procedure} answers YES
whenever $\dF(P,Q)\le r$, and every YES answer comes with a
$Cr$-matching.

\paragraph{Free-space diagrams.}

Alt and Godau~\cite{AG95} introduced the free-space diagram to turn the
decision problem into a reachability problem. The \emph{free-space
diagram} of $P$ and $Q$ is the rectangle $[0,n-1]\times[0,m-1]$, whose
point $(x,y)$ represents the pair of points $P(x)$ and $Q(y)$.
A matching of $P[a,a']$ and $Q[c,c']$ traces a continuous path $\pi$
from $(a,c)$ to $(a',c')$ in this diagram. The path is
\emph{bi-monotone}, meaning that both coordinates are nondecreasing.
Conversely, every such path represents a matching of the two
subcurves. We identify a matching with its path when working in the
diagram.
For a radius $r\ge0$, the \emph{free space} is
$\mathcal F_r=\{(x,y):\|P(x)-Q(y)\|\le r\}$, and its points are
\emph{$r$-free}. Each pair of original edges defines a \emph{cell}
$[i,i+1]\times[j,j+1]$. The free space within a cell is convex. Its
intersection with a cell side is a closed, possibly empty interval,
called a \emph{free interval}.

For points $s=(x,y)$ and $t=(x',y')$ of the diagram, $t$ is
\emph{$r$-reachable} from $s$ if $x\le_P x'$, $y\le_Q y'$, and
$\dF(P[x,x'],Q[y,y'])\le r$. Equivalently, some bi-monotone path in
$\mathcal F_r$ connects $s$ to $t$. In particular,
$\dF(P,Q)\le r$ if and only if $(n-1,m-1)$ is $r$-reachable from
$(0,0)$. Being $r$-free is a property of a single pair of positions,
whereas $r$-reachability also requires the preceding portions of the
curves to be matched within distance $r$
(Figure~\ref{fig:free-space-diagram}(a)). The classical decision
algorithm propagates the reachable parts of the cell sides
from cell to cell. By convexity, the reachable parts of the top and
right sides of a cell are determined by those of its bottom and left
sides in constant time, which decides $\dF(P,Q)\le r$ in $O(nm)$ time.

\input{figures/free-space-diagram}

\paragraph{Reachability propagation.}

To propagate reachability through larger regions, we use rectangles
of the diagram. A \emph{rectangle}
is a set $R=[x_0,x_1]\times[y_0,y_1]$ with
$x_0,x_1\in[0,n-1]$, $y_0,y_1\in[0,m-1]$,
$x_0\le_P x_1$, and $y_0\le_Q y_1$, so its sides
need not lie on cell boundaries (Figure~\ref{fig:free-space-diagram}(b)).
It corresponds to the pair of subcurves
$P[x_0,x_1]$ and $Q[y_0,y_1]$, and its size is
$|R|=|P[x_0,x_1]|+|Q[y_0,y_1]|$. Its \emph{input boundary} $\In(R)$
is the union of its left and bottom sides, and its \emph{output
boundary} $\Out(R)$ is the union of its top and right sides. Their
$r$-free parts are $\In(R)\cap\mathcal F_r$ and
$\Out(R)\cap\mathcal F_r$, shown as thick segments in
Figure~\ref{fig:free-space-diagram}(b). Every bi-monotone path that
crosses $R$ enters through $\In(R)$ and leaves through $\Out(R)$.
The original vertices partition each side into $O(|R|)$
\emph{boundary edges}, and the free part of each boundary edge is a
single free interval. For $s,t\in R$, we write $s\Rr r t$ if some
bi-monotone path from $s$ to $t$ inside $R\cap\mathcal F_r$ exists, and
we also write $\Rr r$ for the set of pairs $(s,t)\in\In(R)\times\Out(R)$
with $s\Rr r t$.

We record reachability information on the boundary of $R$ by a
\emph{labelling} $\lambda$, which assigns a \emph{label} to each
boundary point. The classical dynamic program can be viewed as
propagating Boolean values: $1$ for reachable and $0$ for unreachable.
In our queries, finite labels mark reachable points and can also
distinguish different entrances. Retaining witness information with each propagated label
lets us recover a path from an entrance carrying the same label.
Consecutive points on a boundary edge often
carry the same label, so we represent them by an interval with that
label. These labelled intervals form a \emph{label profile} describing
the information along the edge.

A \emph{reachability query} takes the profiles on $\In(R)$ and computes
profiles on $\Out(R)$, which can serve as input to adjacent rectangles.
Several entrances may reach the same exit, so we need a rule for
selecting the label recorded there. Each call requests one of two
variants: \emph{min-propagation} selects the smallest reachable
entrance label, while \emph{max-propagation} selects the largest.
If no finite label reaches an exit, its value is $+\infty$ for
min-propagation and $-\infty$ for max-propagation. Our approximate
queries relax this selection rule by allowing witnesses at a larger
radius, while satisfying the following guarantee.

\begin{definition}[Certified propagation]
\label{def:certified}
Let $r\ge0$ and $C\ge1$. A labelling $\lambda'$ of $\Out(R)$ is an
\emph{$(r,C)$-certified min-propagation} (respectively,
\emph{max-propagation}) of a labelling $\lambda$ of $\In(R)$ if, for
every $t\in\Out(R)$,
\begin{enumerate}
\item (\emph{completeness}) for every $s\in\In(R)$ with $s\Rr r t$,
\[
 \lambda'(t)\le\lambda(s)\quad\text{for min-propagation},\qquad
 \lambda'(t)\ge\lambda(s)\quad\text{for max-propagation};
\]
\item (\emph{soundness}) if $\lambda'(t)$ is finite, a source
$s\in\In(R)$ and a bi-monotone path $\pi$ from $s$ to $t$ can be
reported with
\[
 \lambda(s)=\lambda'(t),\qquad \pi\subseteq R\cap\mathcal F_{Cr}.
\]
\end{enumerate}
\end{definition}

For a fixed input labelling, certified propagation records one label
at each exit. Completeness ensures that every $r$-reachable entrance
is accounted for when selecting that label. For each finite output
label, soundness gives a \emph{local witness} from an entrance
carrying the same label to the exit, at radius $Cr$. It certifies the
passage through $R$, even if the input label came from an earlier
propagation.

\paragraph{Boundary information.}

For a rectangle $R$, we want to record which entrances in $\In(R)$
can reach which exits in $\Out(R)$. We call this \emph{boundary
information} of $R$.

\begin{definition}[Boundary information]
\label{def:boundary-info}
Let $r\ge0$ and $C\ge1$. A relation
$\mathcal B\subseteq\In(R)\times\Out(R)$ is
\emph{$(r,C)$-boundary information} of $R$ if
\[
 \Rr r\ \subseteq\ \mathcal B\ \subseteq\ \Rr{Cr},
\]
and if, for every pair $(s,t)\in\mathcal B$, a bi-monotone path from
$s$ to $t$ in $R\cap\mathcal F_{Cr}$, called a \emph{witness}, can be
reported.
\end{definition}

The first inclusion ensures that no entrance--exit pair reachable at
radius $r$ is lost. The second guarantees that every recorded pair
can be connected by a path through $R$ at radius $Cr$.
The same boundary information can be reused with different input
labels: the labels specify where the current propagation starts and
what information it carries, while the stored relation determines how
it can pass through $R$.
In Section~\ref{sec:boundary}, we construct boundary information from
a few certified propagations with suitably chosen labels, using the
geometry of free space. We also show how to combine and apply this
information efficiently, with time bounds for these operations.

\paragraph{Preprocessing and queries.}

Our algorithm uses one preprocessing routine and two query routines.
Given a curve $H$ and a radius $r$, $\Build(H,r)$ stores boundary
information and witness data for selected rectangles in $H\times H$.
We call $H$ the \emph{host} and the resulting data structure a
\emph{self-index}. The stored information is independent of the
labels supplied by later queries.

Given this index, $\SelfQuery(U,V,\lambda)$ takes two subcurves $U,V$
of $H$ and input labels $\lambda$ on $\In(U\times V)$. It uses the
stored information to compute output label profiles on
$\Out(U\times V)$, together with information for reporting local
witnesses.

The general query $\Query(P',Q',\lambda)$ takes subcurves $P',Q'$ of
two separately preprocessed curves $P,Q$, together with input labels
$\lambda$ on $\In(P'\times Q')$. It returns the same kind of output
on $\Out(P'\times Q')$, using the separate indexes without a joint
index on $P\times Q$.

Both query routines provide certified propagation: every exit that
is $r$-reachable from a finite-labelled entrance receives a finite
label, and each finite output has a local witness from an entrance
carrying the same label at the allowed larger radius.

%% file: figures/free-space-diagram.tex
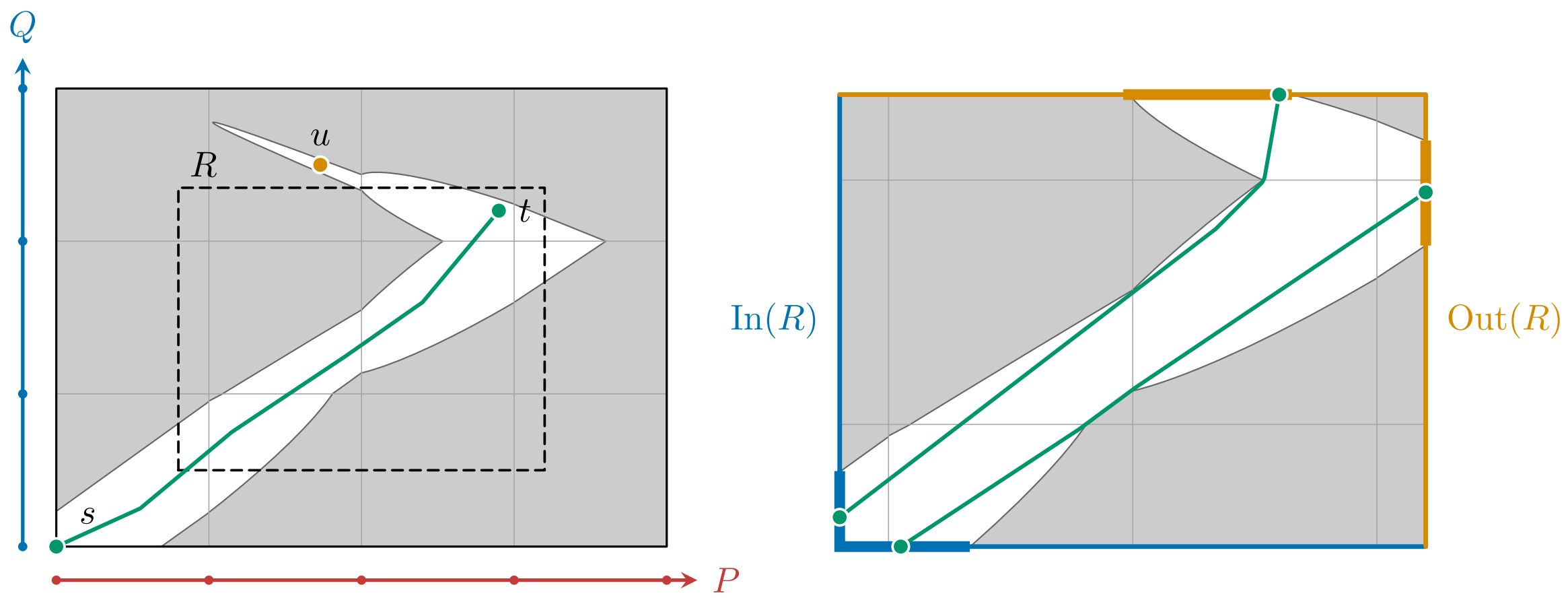
\begin{figure}[t]
\centering
\begin{tikzpicture}[>=stealth,line cap=round,line join=round,
  every node/.style={font=\small},
  forbidden/.style={fill=black!20},
  freeedge/.style={draw=black!60,line width=.4pt},
  cellgrid/.style={draw=black!35,line width=.3pt},
  curveaxis/.style={->,line width=1pt},
  wit/.style={line width=1.1pt,figgreen},
  bar/.style={line width=3pt,line cap=butt},
  pt/.style={circle,inner sep=0pt,minimum size=4.6pt,draw=white,
    line width=.7pt}]
\def\FScell#1#2#3#4#5#6#7{%
 \begin{scope}
  \clip (#1,#2) rectangle ++(1,1);
  \filldraw[fill=white,freeedge]
   (#3,#4) circle [x radius=#5,y radius=#6,rotate=#7];
 \end{scope}}
\def\FSregions{%
 \FScell{0}{0}{25.333}{18.000}{74.479}{0.323}{-144.41}%
 \FScell{1}{0}{0.407}{0.195}{1.787}{0.329}{-145.10}%
 \FScell{1}{1}{-4.792}{-3.125}{9.836}{0.306}{-146.62}%
 \FScell{1}{2}{36.778}{-11.667}{38.560}{0.208}{-22.00}%
 \FScell{2}{1}{3.190}{2.080}{1.704}{0.310}{-147.02}%
 \FScell{2}{2}{3.168}{1.959}{1.280}{0.211}{-20.63}%
 \FScell{3}{1}{1.917}{1.250}{9.836}{0.306}{-146.62}%
 \FScell{3}{2}{8.000}{0.000}{38.560}{0.208}{-22.00}%
}
\begin{scope}[x=1.5cm,y=1.5cm]
 \fill[forbidden] (0,0) rectangle (4,3);
 \FSregions
 \foreach \x in {1,2,3} \draw[cellgrid] (\x,0)--(\x,3);
 \foreach \y in {1,2} \draw[cellgrid] (0,\y)--(4,\y);
 \draw[line width=.6pt] (0,0) rectangle (4,3);
 \draw[curveaxis,figred] (0,-.22)--(4.2,-.22) node[right] {$P$};
 \draw[curveaxis,figblue] (-.22,0)--(-.22,3.2) node[above] {$Q$};
 \foreach \x in {0,...,4} \fill[figred] (\x,-.22) circle[radius=1.3pt];
 \foreach \y in {0,...,3} \fill[figblue] (-.22,\y) circle[radius=1.3pt];
 \draw[densely dashed,line width=.7pt] (0.8,0.5) rectangle (3.2,2.35);
 \node[above right=-1pt] at (0.8,2.35) {$R$};
 \draw[wit] (0,0)--(0.55,0.25)--(1.15,0.75)--(1.9,1.25)--(2.4,1.6)
  --(2.9,2.2);
 \node[pt,fill=figgreen,label={[label distance=1pt]above right:$s$}]
  at (0,0) {};
 \node[pt,fill=figgreen,label={[label distance=-1pt]right:$t$}]
  at (2.9,2.2) {};
 \node[pt,fill=figorange,label={[label distance=-1pt]above:$u$}]
  at (1.73,2.5) {};
 \node at (2,-.68) {(a) Free space};
\end{scope}
\begin{scope}[xshift=7.7cm,x=2.4cm,y=2.4cm]
 \begin{scope}[shift={(-0.8,-0.5)}]
  \begin{scope}
   \clip (0.8,0.5) rectangle (3.2,2.35);
   \fill[forbidden] (0.8,0.5) rectangle (3.2,2.35);
   \FSregions
   \foreach \x in {1,2,3} \draw[cellgrid] (\x,0.5)--(\x,2.35);
   \foreach \y in {1,2} \draw[cellgrid] (0.8,\y)--(3.2,\y);
  \end{scope}
  \draw[figblue,line width=1.3pt] (0.8,2.35)--(0.8,0.5)--(3.2,0.5);
  \draw[figorange,line width=1.3pt] (0.8,2.35)--(3.2,2.35)--(3.2,0.5);
  \draw[bar,figblue,line join=miter] (0.8,0.809)--(0.8,0.5)--(1.333,0.5);
  \draw[bar,figorange] (3.2,1.732)--(3.2,2.162);
  \draw[bar,figorange] (1.961,2.35)--(2.652,2.35);
  \draw[wit] (1.05,0.5)--(1.78,0.98)--(2.01,1.15)--(3.2,1.95);
  \draw[wit] (0.8,0.62)--(2.34,1.8)--(2.53,1.99)--(2.54,2.01)--(2.6,2.35);
  \foreach \p in {(1.05,0.5),(0.8,0.62),(3.2,1.95),(2.6,2.35)}
   \node[pt,fill=figgreen] at \p {};
  \node[figblue,left=2pt] at (0.8,1.425) {$\In(R)$};
  \node[figorange,right=2pt] at (3.2,1.425) {$\Out(R)$};
  \node at (2,.075) {(b) Boundary reachability};
 \end{scope}
\end{scope}
\end{tikzpicture}
\caption{(a)~Free space at radius $r$ (white). Both $t$ and $u$ are
$r$-free, but only $t$ is $r$-reachable from $s$.
(b)~Input and output boundaries of $R$ (blue and orange); thick
segments are $r$-free. Green paths witness reachability.}
\label{fig:free-space-diagram}
\end{figure}

%% file: sections/03-technique-overview.tex

\section{Algorithm overview}
\label{sec:technique-overview}

We focus on describing the decision procedure for the Fr\'echet
distance at a high level. The ideas can be adapted to the discrete
Fr\'echet distance.

\subsection{Preprocessing and self-queries}
\label{ix:base}

Let $\Build_i$ and $\SelfQuery_i$ denote the routines in round $i$.
For a host curve $H$ of size $b$, $\Build_i(H,r)$ prepares boundary
information for a family of rectangles in $H\times H$. We call the
resulting self-index simply an \emph{index}. Given subcurves $U,V$
of $H$, $\SelfQuery_i(U,V,\lambda)$ takes input labels $\lambda$
describing the reachability information arriving at the entrances of
$U\times V$. It uses the stored boundary information to pass this
information to the exits, where it is again represented by labels.

\paragraph{What the index stores.}
We illustrate the stored family with the base routine $\Build_0$.
Organize the edges of $H$ into a dyadic hierarchy: at each scale,
group consecutive edges into aligned intervals containing
$1,2,4,\ldots$ edges. We call these \emph{canonical intervals}. For every
ordered pair of canonical intervals $I,J$, $\Build_0$ stores a
\emph{boundary map}, a representation of the boundary information of
the \emph{canonical rectangle} $I\times J$. The two
intervals may come from different scales, so the index includes long,
thin rectangles as well as squares. Each map records which entrances
can reach which exits, up to a constant approximation, and retains
the information needed to report matching witnesses. It can then be
applied to new input labels without recomputing reachability inside
the rectangle. Figure~\ref{fig:base-index}(a) shows the interval
hierarchy.

The base routine $\Build_0$ constructs these maps using cell-by-cell
propagation. At any fixed pair of scales, the canonical rectangles
partition $H\times H$ and contain $O(b^2)$ cells in total. Summing over
the $O(\log^2 b)$ pairs of scales gives $\tO(b^2)$ preprocessing
time. Here and below, $\tO$ suppresses polylogarithmic factors.

\input{figures/base-index}

\paragraph{Using the stored maps.}
The routine $\SelfQuery_i$ partitions the query rectangle into
canonical rectangles, whose maps are stored, and boundary pieces
whose maps are computed on demand. For the base routine
$\SelfQuery_0$, split each of $U,V$ into $O(\log b)$ disjoint
canonical intervals and at most two partial original edges at its
ends, choosing the largest canonical intervals that fit. Products of
the canonical intervals give $O(\log^2 b)$ rectangles whose maps are
already stored. The remaining thin end strips are handled by exact
propagation. Figure~\ref{fig:base-index}(b) shows such a query
partition.

$\SelfQuery_0$ combines these maps into a map of $U\times V$ and
applies it to the input labels. During the combination, rectangles
are processed from bottom-left to top-right, passing labels across
shared sides. Each stored rectangle is processed
in time near-linear in its boundary size. Since the intervals on
each axis are disjoint, the canonical products have total boundary
size $O(b\log b)$, and the thin strips meet only $O(b)$ cells.
Thus $\SelfQuery_0$ takes $\tO(b)$ time, even when the queried
rectangle contains quadratically many cells. The resulting queries
have a constant approximation factor. Section~\ref{ix:section}
gives the construction and its guarantees.

We will replace this pair of algorithms by successively cheaper
preprocessing algorithms, while keeping self-queries fast. We first use
$\SelfQuery_i$ to speed up queries between two separately prepared
curves. For $i\ge1$, the general queries from round $i-1$ are then used to
construct $\Build_i$ and $\SelfQuery_i$.

\subsection{The inner recursion: from self-queries to general queries}
\label{intro:surrogates}

A general query computes boundary reachability between two separately
preprocessed curves $X,Y$, each of size at most $b$. Given labels on
the input boundary $\In(X\times Y)$, it propagates them to the output
boundary $\Out(X\times Y)$. The two indices store information about
subcurve pairs within $X$ and within $Y$, but no boundary map for
$X\times Y$ has been stored.

We start with the base indices and describe $\Query_{0,1}$ using the
exact cell-by-cell routine $\Query_{0,0}$. There are three steps.
We describe sources on the bottom side. Sources on the left side are
handled by exchanging the axes, and the two answers are combined.

\paragraph{1. Partition $Y$.}
Cut $Y$ into $k$ consecutive chunks $Y_1,\ldots,Y_k$, each of size
$O(q)$, where $q=\lceil b/k\rceil$. We will replace these chunks by
matching subcurves of $X$. This turns the main part of the general
query into self-queries on $X$.

\paragraph{2. Find matching subcurves by a cursor search.}
Maintain a cursor on $X$, initially at its start. For the current
chunk $Y_j$, test prefixes of the remaining part of $X$ of lengths
$1,2,4,\ldots$ using $\Query_{0,0}$, truncating at the endpoint of
$X$ if necessary. Stop at the first successful test or after testing
the whole remaining part of $X$. Each test starts from all free
points on the bottom of its rectangle and asks whether any top point
is reachable. At the first successful test, take a witness to the
leftmost reachable top point. Its projection onto $X$ gives a
\emph{donor} $U_j$ matching $Y_j$.

For the next chunk, start the search at the endpoint of $U_j$ on $X$,
so successive donors appear in order along $X$.
A full donor need not exist, even for the first chunk $Y_1$. If no
test succeeds by the time the search reaches the end of $X$, we stop.
A matching may still reach the end of $X$ before finishing $Y_j$,
giving an output on the right boundary. If so, we take a witness to
the highest reachable point there. Its subcurve of $X$ is a
\emph{partial donor}: it matches only a prefix of $Y_j$.
The next step uses the donors found so far, including this final
partial donor if one exists. If $Y_1$ has neither a full nor a partial
donor, no output is reachable from the bottom side. Sources on the
left side are still handled separately.

\paragraph{3. Answer self-queries on the donors.}
We form a \emph{surrogate} $S$ by keeping the donors in their order
along $X$ and connecting consecutive donors by straight segments.
These \emph{joins} shortcut the gaps between donors. Both ends of
each join are within distance $r$ of the common
endpoint of the adjacent chunks, so the whole segment is as well.
The saved donor matchings therefore extend to an $r$-matching $\mu$
between $S$ and the prefix of $Y$ represented by the search.

We next propagate the reachability information carried by the input
labels through $X\times S$ at radius $2r$, preserving all
$r$-reachability in $X\times Y$ by the triangle inequality with $\mu$.
Each rectangle $X\times U_j$ pairs the whole of $X$ with one of its
subcurves. $\SelfQuery_0$ combines the stored canonical maps to
propagate all input labels together in $\tO(b)$ time,
without scanning the $O(b|U_j|)$ cells.
We process these rectangles in order and collect their right outputs.
Across a join, let $X$ pause while $S$ traverses the joining segment.
We pass the top-output labels to the same positions on the
next bottom boundary wherever both join endpoints are within $2r$
of the point on $X$. Convexity makes this traversal valid. Every
original $r$-matching is preserved: when $Y$ reaches the common
chunk endpoint $p$, its matched point on $X$ and the whole join
are within $r$ of $p$, so the traversal has cost at most $2r$.
Each join is handled by a scan along $X$. Together, the self-queries
and joins take $\tO(kb)$ time for all bottom sources
(Figure~\ref{fig:shared-surrogate}). We concatenate the local
witnesses, and $\mu$ carries the outputs back to $X\times Y$. Using the stored boundary information, we can perform
this mapping in $\tO(b)$ time (Lemma~\ref{qr:records}).

\input{figures/shared-surrogate}

The three steps above, repeated with the axes exchanged, define
$\Query_{0,1}$. Its running time has two main contributions: the
cursor search and the self-queries.
For the cursor search, a $\Query_{0,0}$ test of width $w$ against
$Y_j$ costs $\tO(wq)$ time. For each successful chunk, let $w_j$ be
the increase in the cursor's $X$-coordinate and $\ell_j$ the number
of failed tests. The first successful test has width at most
$2^{\ell_j}$. Exactness gives $2^{\ell_j-1}<w_j$ for $\ell_j\ge1$,
so the total tested width is at most
\[
 \underbrace{\sum_{h=0}^{\ell_j-1}2^h}_{\text{failed tests}}
 +\underbrace{2^{\ell_j}}_{\text{first success}}
 =2^{\ell_j+1}-1=O(1+w_j).
\]
For $\ell_j=0$, this is just the initial width-$1$ test, even if
$w_j=0$. Hence the search for $Y_j$ costs $\tO(q(1+w_j))$.
Summing over successful chunks gives $\sum_jw_j\le b$, and a final
unsuccessful search, if any, costs $\tO(bq)$. The total search time is
$\tO(q\sum_j(1+w_j)+bq)=\tO(bq)=\tO(b^2/k)$.
The second part makes at most $k$ calls to $\SelfQuery_0$, each
taking $\tO(b)$ time, for a total of $\tO(kb)$. Repeating the procedure
with the axes exchanged changes the cost by only a constant factor,
so $\Query_{0,1}$ takes $\tO(b^2/k+kb)$ time.
Choosing $k\asymp\sqrt b$ balances the two terms and gives
$\tO(b^{3/2})$ time.
The preprocessing on $X$ and $Y$ is reused throughout. Since
$\SelfQuery_0$ has a constant approximation factor, the triangle
inequality through the $r$-matching $\mu$ gives a constant factor for
$\Query_{0,1}$ as well.

\paragraph{The recursive query.}
We now describe $\Query_{i,t}$, using the preprocessing routine
$\Build_i$ and the self-query routine $\SelfQuery_i$. Given separately
preprocessed $X,Y$, we follow the same three steps: partition $Y$
into $k$ chunks, use cursor search to construct a surrogate $S$, and
apply $\SelfQuery_i$ to propagate the input labels through
$X\times S$. For $t\ge1$, each search test calls $\Query_{i,t-1}$.
The base case $\Query_{i,0}$ uses exact cell-by-cell propagation.

The key observation is that each search test is itself a general
query with one short side, so it can use the preceding level of the
recursion. Faster tests let us use fewer chunks and hence fewer
self-queries. The recursion reduces the depth $t$, while reusing the
round-$i$ indexes and $\SelfQuery_i$.

Suppose the given $\SelfQuery_i$ routine takes
$\tO(b^{1+\delta_i})$ time (we establish this bound later).
We analyze $\Query_{i,t}$ by induction
on its recursion depth $t$, deriving a total running-time bound of the form
$\tO(b^{1+\beta_{i,t}})$. The exact base case takes $\tO(b^2)$ time,
giving $\beta_{i,0}=1$. Let $A_{i,t}$ be the query's approximation
factor. Algorithm~\ref{alg:query-overview} gives the framework and
the resulting time bounds.
The routine $\mathsf{CursorSearch}_{i,t-1}$ partitions $Y$ into $k$
chunks and uses $\Query_{i,t-1}$ to construct $S$ and $\mu$, with
$y^*$ denoting the endpoint of the represented prefix of $Y$.
The routine $\mathsf{SurrogateQuery}_i$ then propagates the input
labels through the donors using $\SelfQuery_i$ and transfers them
across the joins. Section~\ref{sec:query} gives these components.
The chunk size is $O(q)$, where $q=\lceil b/k\rceil$. We choose $k$
in the analysis below.

\begin{algorithm}[!htb]
\caption{$\Query_{i,t}(X,Y,\lambda)$: the recursive framework}
\label{alg:query-overview}
\begin{algorithmic}[1]
\Require Separate round-$i$ indexes for $X,Y$, depth $t\ge0$, radius $r$, and input labels $\lambda$
\Ensure Propagated output labels and matching witnesses
\If{$t=0$}
 \State \Return exact cell-by-cell propagation of $\lambda$ through $X\times Y$
\EndIf
\For{each input side with labels $\lambda_{\rm side}$ (exchange axes for the left side)}
 \State $(S,\mu,y^*)\gets\mathsf{CursorSearch}_{i,t-1}(X,Y,k,r)$
 \Statex \Comment{$\tO(bq^{\beta_{i,t-1}})$ total, Lemma~\ref{qr:cursor}}
 \State If $S$ is empty, record an empty answer for this side and continue
 \State $\eta\gets\mathsf{SurrogateQuery}_i(X,S,\lambda_{\rm side},(A_{i,t-1}+1)r)$
 \Statex \Comment{$\tO(kb^{1+\delta_i})$ total, Lemma~\ref{qr:transfer}}
 \State Map $\eta$ back to $X\times Y$ through $\mu$, retaining top outputs only if $y^*$ reaches the end of $Y$
 \Statex \Comment{$\tO(b)$, Lemma~\ref{qr:records}}
\EndFor
\State Combine the output labels from the two input sides and \Return the result
\end{algorithmic}
\end{algorithm}

\paragraph{Running time.}
As in the analysis of $\Query_{0,1}$, the main costs are the cursor
search and the self-queries. The difference is that the search now
calls $\Query_{i,t-1}$ instead of exact propagation.
The cursor moves at least to the end of the last failed prefix, if
there is one. If a search succeeds after a failure, it therefore
advances by at least half the largest tested width. By doubling, the
sum of all tested widths is $O(b+k)=O(b)$, including unit-width
searches and a possible final unsuccessful search. A width-$W$ test
against a chunk of size $O(q)$ costs
$\tO((W+q)q^{\beta_{i,t-1}})$. Since each chunk uses $O(\log b)$
tests and $kq=O(b)$, the total search time is
$\tO(bq^{\beta_{i,t-1}})$. Lemma~\ref{qr:cursor} gives the detailed
analysis.

There are at most $k$ calls to $\SelfQuery_i$, each taking
$\tO(b^{1+\delta_i})$ time. Each join requires one scan along $X$,
costing $\tO(b)$, so the self-queries and joins together take
$\tO(kb^{1+\delta_i})$ time (Lemma~\ref{qr:transfer}). Mapping the
outputs back takes $\tO(b)$ time (Lemma~\ref{qr:records}). Thus,
including both input sides, the total query time is
\begin{equation}
 \tO\bigl(b(b/k)^{\beta_{i,t-1}}+kb^{1+\delta_i}\bigr).
 \label{eq:overview-query-cost}
\end{equation}
For $0\le\delta_i<1$, choosing
$k\asymp b^{(\beta_{i,t-1}-\delta_i)/(1+\beta_{i,t-1})}$ balances the two
terms and gives the recurrence
\begin{equation}
 \beta_{i,t}=\frac{\beta_{i,t-1}(1+\delta_i)}{1+\beta_{i,t-1}}.
 \label{eq:overview-inner-time}
\end{equation}
Starting from $\beta_{i,0}=1$, this gives $\beta_{i,t}=1/(t+1)$ when
$\delta_i=0$, and
$\beta_{i,t}=\delta_i/(1-(1-\delta_i)(1+\delta_i)^{-t})$ otherwise.
Thus, by choosing a sufficiently large recursion depth $t$, we can
make the general-query running time arbitrarily close in the exponent
to the self-query running time.

\paragraph{Approximation factor.}
Let $D_i$ be the approximation factor of $\SelfQuery_i$. Exact
propagation gives $A_{i,0}=1$. The recursive search supplies an
$A_{i,t-1}r$-matching between $S$ and the represented prefix of $Y$.
An original $r$-matching therefore gives a matching in $X\times S$
of cost at most $(A_{i,t-1}+1)r$. We run $\SelfQuery_i$ at this enlarged
radius, using indexes prepared at the required radii. Its witnesses
have cost at most $D_i(A_{i,t-1}+1)r$. Mapping back through $\mu$ adds
$A_{i,t-1}r$.
Thus
\begin{equation}
 A_{i,t}=D_i(A_{i,t-1}+1)+A_{i,t-1},\qquad
 A_{i,t}=2(D_i+1)^t-1.
 \label{eq:overview-inner-factor}
\end{equation}
The number of donors does not enter this factor: concatenating their
witnesses takes the maximum of their costs. Sections~\ref{sec:boundary}
and~\ref{sec:query} give the boundary operations and the full analysis.

\subsection{The outer recursion: cheaper preprocessing}
\label{intro:index}

The base routine $\Build_0$ still takes $\tO(b^2)$ time. Its
self-queries take $\tO(b)$ time, and a sufficiently large depth $T$
brings the running time of $\Query_{0,T}$ close to linear, but the
preprocessing remains quadratic. The outer recursion reduces this
cost by preparing the old indexes only on short blocks of the host.
Fast general queries then supply the boundary information between
blocks, from which we build the new index.
For $i\ge1$, suppose inductively that $\Build_{i-1}$ and
$\SelfQuery_{i-1}$ take $\tO(b^{1+\alpha_{i-1}})$ and
$\tO(b^{1+\delta_{i-1}})$ total time, respectively. The base routines
satisfy these bounds with $\alpha_0=1$ and $\delta_0=0$.
Using $\Build_{i-1}$ and $\Query_{i-1,T}$, we construct $\Build_i$
and $\SelfQuery_i$ and derive the new excess exponents
$\alpha_i$ and $\delta_i$ below.
In Algorithms~\ref{alg:build-overview} and~\ref{alg:selfquery-overview},
we choose $T$ so that
$\operatorname{time}(\Query_{i-1,T})\le\tO(b^{1+\theta})$
on curves of size at most $b$. We specify $\theta$ and $T$ in the
running-time analysis below. The construction of $\Build_i$ has
three steps.

\paragraph{1. Partition the host.}
Cut the host $H$ of size $b$ into a family $\mathcal B$ of
$O(b/h)$ consecutive blocks, each of size $O(h)$.

\paragraph{2. Prepare the blocks.}
Use $\Build_{i-1}$ to prepare self-indexes on each block and on
selected shorter subcurves within it, at the scales and radii needed
by $\Query_{i-1,T}$. By induction, each index supports
$\SelfQuery_{i-1}$ on arbitrary subcurves of its host, and its time
bound already accounts for further recursion. The shorter hosts let
the inner recursion use indexes at the appropriate local scale.
Collect these indexes in $\mathcal J$.

\paragraph{3. Store the boundary maps.}
Using $\mathcal J$, we first construct a \emph{primitive boundary
map} for each block-pair rectangle with $\Query_{i-1,T}$.
The canonical rectangles have sides consisting of aligned groups of
$1,2,4,\ldots$ blocks. Each is therefore a grid of rectangles whose
primitive maps are already available. We use these maps to propagate
boundary information across the grid, passing outputs of one rectangle
to inputs of its neighbors, and store the resulting entrance--exit
information as a map of the whole rectangle. The routine
$\mathsf{StoreMaps}_{i-1,T}$ collects these maps, over all pairs of
scales, in $\mathcal M$.

Algorithm~\ref{alg:build-overview} summarizes the construction.
Section~\ref{sec:index} gives the implementation of $\mathsf{StoreMaps}$
and proves the bounds in Lemma~\ref{ix:tradeoff}.

\begin{algorithm}[!htb]
\caption{$\Build_i(H,r)$: the outer recursive step}
\label{alg:build-overview}
\begin{algorithmic}[1]
\Require The level-$(i-1)$ routines, depth $T$, target $\theta$, and a host $H$ of size $b$
\Ensure A level-$i$ self-index
\State Set $h\asymp b^{1/(1+\alpha_{i-1}-\theta)}$
\State Partition $H$ into blocks $\mathcal B$ of size $O(h)$
\Statex \Comment{$\tO(b)$}
\For{each block $B\in\mathcal B$}
 \State Use $\Build_{i-1}$ to prepare the required indexes on $B$ and add them to $\mathcal J$
\EndFor
\Statex \Comment{$\tO(bh^{\alpha_{i-1}})$, Lemma~\ref{ix:bundles}}
\State $\mathcal M\gets\mathsf{StoreMaps}_{i-1,T}(\mathcal B,\mathcal J,r)$
\Statex \Comment{$\tO(b^2h^{\theta-1})$, Lemma~\ref{ix:tradeoff}}
\State \Return the blocks $\mathcal B$, block indexes $\mathcal J$, and maps $\mathcal M$
\end{algorithmic}
\end{algorithm}

\paragraph{Self-queries in the new index.}
Given subcurves $U,V$ of $H$, we must propagate the input labels
through $R=U\times V$. First separate the whole-block middle of each
subcurve from the partial blocks at its ends. Each middle is a union
of $O(\log b)$ canonical intervals, so their product is covered by
$O(\log^2 b)$ canonical rectangles $\mathcal C$.
Cut the remaining strips and corners along
block boundaries into $O(b/h)$ \emph{fringe rectangles}
$\mathcal F$. Each fringe rectangle lies inside a single pair of
prepared blocks.

We retrieve the stored map of each rectangle in $\mathcal C$.
For a fringe rectangle $F=U_F\times V_F$, we apply
$\Query_{i-1,T}$ to the subcurves $U_F,V_F$, using the indexes of
their containing blocks from Step~2. A few calls with suitably
chosen labels construct the boundary map of $F$. Let $\mathcal M_R$
be the resulting collection of local maps.

These maps form a network: outputs from one rectangle become inputs
to its neighbors. By propagating suitable labels through this network,
we construct a single boundary map $B_R$ of $R$. We then apply the
caller's labels $\lambda$ to $B_R$, obtaining output profiles and
local witness records. This last step is the only one that uses
$\lambda$. The map construction is independent of it.
Algorithm~\ref{alg:selfquery-overview} gives these steps, with the
details in Section~\ref{sec:index}.

\begin{algorithm}[!htb]
\caption{$\SelfQuery_i(U,V,\lambda)$}
\label{alg:selfquery-overview}
\begin{algorithmic}[1]
\Require The index $\mathcal I$ from $\Build_i(H,r)$, subcurves $U,V$, and input labels $\lambda$
\Ensure Propagated output profiles and local witness records
\State Decompose $R=U\times V$ into canonical rectangles $\mathcal C$ and fringes $\mathcal F$
\Statex \Comment{$\tO(b/h)$, Lemma~\ref{ix:tradeoff}}
\State $\mathcal M_R\gets$ the stored maps of the rectangles in $\mathcal C$
\For{each fringe rectangle $F\in\mathcal F$}
 \State Add a map of $F$, compiled using $\Query_{i-1,T}$, to $\mathcal M_R$
\EndFor
\Statex \Comment{$\tO(bh^\theta)$, Lemma~\ref{ix:tradeoff}}
\State Compile network propagation through $\mathcal M_R$ into a boundary map $B_R$ of $R$
\Statex \Comment{$\tO(b)$, Lemma~\ref{ix:compiler}}
\State \Return the propagation of $\lambda$ through $B_R$
\Statex \Comment{$\tO(b)$, Lemma~\ref{fd:profiles}}
\end{algorithmic}
\end{algorithm}

\paragraph{Running time.}
Let $K$ be the number of outer rounds. The parameter $\theta$ sets
the target time $\tO(h^{1+\theta})$ for general queries on blocks of
size $O(h)$. To obtain the final near-linear bound, we choose
$K=\Theta(1/\eps)$ and $\theta=\Theta(\eps)$.
Lemma~\ref{ix:parameters} gives a common inner depth
$T=O(\theta^{-1}\log(2/\theta))$ for which
$\beta_{i-1,T}\le\theta<\alpha_{i-1}$ for every $1\le i\le K$,
so $\Query_{i-1,T}$ meets this target in each round.

Preparing the $O(b/h)$ blocks costs
$\tO((b/h)h^{1+\alpha_{i-1}})=\tO(bh^{\alpha_{i-1}})$.
There are $O((b/h)^2)$ block pairs, so constructing their primitive
maps at $\tO(h^{1+\theta})$ time per pair costs
$\tO(b^2h^{\theta-1})$ in total. Combining them into canonical maps
adds $\tO(b^2/h)$, which is absorbed by this bound
(Lemma~\ref{ix:tradeoff}). Thus $\Build_i$ takes
$\tO(bh^{\alpha_{i-1}}+b^2h^{\theta-1})$ time.
For $\SelfQuery_i$, the $O(b/h)$ fringe rectangles each cost
$\tO(h^{1+\theta})$ time. The remaining boundary operations take
$\tO(b)$ time, so the total is $\tO(bh^\theta)$, including the report
of one local witness (Lemma~\ref{ix:tradeoff}).
Balancing the two preprocessing terms gives
$h\asymp b^{1/(1+\alpha_{i-1}-\theta)}$, so
$\alpha_i=\alpha_{i-1}/(1+\alpha_{i-1}-\theta)$ and
$\delta_i=\theta/(1+\alpha_{i-1}-\theta)$.
With $\alpha_0=1$ and $\delta_0=0$, these recurrences give
\begin{equation}
 \alpha_i=\frac{\theta}{1-(1-\theta)^{i+1}},\qquad
 \delta_i=\frac{\theta\bigl(1-(1-\theta)^i\bigr)}
                    {1-(1-\theta)^{i+1}}.
 \label{eq:overview-outer-closed}
\end{equation}
The preprocessing exponent decreases towards $1+\theta$, while
the self-query exponent stays below $1+\theta$.

\paragraph{Approximation factor.}
Combining maps of adjacent rectangles concatenates their witnesses
and introduces no loss. Compiling a propagation routine into a
boundary map adds $4$ to its factor to handle arbitrary boundary
points (Section~\ref{fd:normalization}). We compile at three places:
the primitive maps, the canonical maps, and the whole self-query.
Consequently,
\begin{equation}
 D_i=A_{i-1,T}+12=2(D_{i-1}+1)^T+11.
 \label{eq:overview-outer-factor}
\end{equation}
The loss depends on the recursion depths, but not on the number of
blocks or canonical scales.

\subsection{Completing the algorithm}
\label{sec:schedule}
\label{sec:decide-overview}

Fix $0<\eps\le1$. We take $K=\lceil6/\eps\rceil$ outer rounds and
set $\theta=1/K$, targeting general-query time $\tO(b^{1+\theta})$
on curves of size at most $b$. We use the same inner depth
$T=\lceil(4/\theta)\log(16/\theta)\rceil$ in every round.
These parameters depend only on $\eps$, and Lemma~\ref{ix:parameters}
ensures $\beta_{i-1,T}\le\theta<\alpha_{i-1}$ for every $1\le i\le K$.

Since $(1-\theta)^{K+1}<e^{-1}$,
\eqref{eq:overview-outer-closed} gives
$\alpha_K<\theta/(1-e^{-1})<\eps/2$ and
$\delta_K<\theta<\eps/4$. Thus $\Build_K$ on a host of size $b$
takes $\tO(b^{1+\eps/2})$ time, and $\SelfQuery_K$ takes
$\tO(b^{1+\eps/4})$ time. These bounds include all recursive work:
$\Build_K$ constructs the lower-round indexes on shorter hosts.
For a threshold $r$, join $P$ and $Q$ by a segment to form a host
$H$ of size $O(n)$. After $\Build_K(H,r)$, one $\SelfQuery_K$
tests reachability from the initial to the final corner of
$P\times Q$ (Algorithm~\ref{alg:decide}). The query stays inside
this rectangle, so the joining segment is not used.

\begin{algorithm}[!htb]
\caption{$\Decide(P,Q,r)$}
\label{alg:decide}
\begin{algorithmic}[1]
\Require Curves $P,Q$, radius $r>0$, and the parameters chosen above
\Ensure YES with a $D_Kr$-matching, or NO certifying $\dF(P,Q)>r$
\State Join $P$ and $Q$ by a segment to form a host $H$
\State $\mathcal I\gets\Build_K(H,r)$ \Comment{$\tO(n^{1+\eps/2})$}
\State Set $\lambda=0$ at the initial corner of $P\times Q$ and $+\infty$ on the remaining input boundary
\State $\lambda'\gets\SelfQuery_K(P,Q,\lambda)$ using $\mathcal I$ \Comment{$\tO(n^{1+\eps/4})$}
\If{the final corner has a finite label in $\lambda'$}
 \State \Return YES and the corresponding matching
\Else
 \State \Return NO
\EndIf
\end{algorithmic}
\end{algorithm}

Certified propagation answers YES whenever $\dF(P,Q)\le r$,
and every YES answer comes with a $D_Kr$-matching. The total time,
including preprocessing and reporting the matching, is
$\tO(n^{1+\eps/2})$.
To bound $D_K$, recall that each outer round uses $T$ levels of the
inner recursion, giving the power $(D_{i-1}+1)^T$
in~\eqref{eq:overview-outer-factor}. Constructing the boundary maps
adds only a constant to the factor. With $E_i=D_i+1$, we obtain
$E_i\le4E_{i-1}^T$. Taking logarithms gives
$\log E_i\le(T+1)\log E_{i-1}$, so the $K$ rounds starting from
constant $E_0$ yield
$\log E_K=O((T+1)^K)$ and
$\log\log E_K=O(K\log(T+1))=O(\eps^{-1}\log(2/\eps))$.

To approximate the distance itself, we search over the radius using
Lemma~\ref{ix:search}. An $O(n)$-factor estimate, computed in linear
time, restricts the search to $O(\log n)$ radii spaced by factors of
two. Binary search using the decision certificates takes
$O(\log\log n)$ calls. The spacing introduces an extra factor of
two, so the cost of the best retained matching is a
$2D_K$-approximation. Thus
$f(1/\eps)=2D_K\le\exp(\exp(O(\eps^{-1}\log(2/\eps))))$.
Absorbing the search overhead and hidden polylogarithmic factors
into $n^{\eps/2}$ gives $O(n^{1+\eps})$ total time and proves
Theorem~\ref{thm:main}.

%% file: figures/base-index.tex
\begin{figure}[!htb]
\centering
\begin{tikzpicture}[>=stealth,line cap=round,line join=round,
  every node/.style={font=\small},
  interval/.style={draw=figblue!65,fill=figblue!8,line width=.7pt},
  chosen/.style={draw=figgreen,fill=figgreen!18,line width=1.1pt},
  map/.style={draw=figblue,line width=1pt},
  grid/.style={draw=black!20,line width=.35pt}]
\begin{scope}[x=.34cm,y=1cm]
 \foreach \len/\lev/\last in {1/.65/15,2/1.35/7,4/2.05/3,8/2.75/1,16/3.45/0}{
  \node[left=5pt,black!65] at (0,\lev) {$\len$};
  \foreach \j in {0,...,\last}{
   \draw[interval] ({\j*\len+.08},\lev-.075)
    rectangle ({(\j+1)*\len-.08},\lev+.075);
  }
 }
 \node[black!65] at (-.85,3.9) {edges};
 \draw[chosen] (4.08,1.975) rectangle (7.92,2.125);
 \node[above=3pt,figgreen] at (6,2.125) {$I$};
 \draw[chosen] (8.08,1.275) rectangle (9.92,1.425);
 \node[above=3pt,figgreen] at (9,1.425) {$J$};
 \draw[->,line width=.8pt] (0,0)--(16.6,0) node[right] {$H$};
 \foreach \j in {0,...,16}
  \fill[black!70] (\j,0) circle[radius=1.1pt];
 \node at (8,-.75) {(a) Canonical intervals at all scales};
\end{scope}
\begin{scope}[shift={(7.7,.05)},x=.48cm,y=.70cm]
 \fill[figorange!22] (0,0) rectangle (11,5);
 \fill[figblue!7] (.5,.5) rectangle (10.5,4.5);
 \fill[figgreen!17] (1.5,2.5) rectangle (5.5,4.5);
 \foreach \x in {.5,1.5,...,10.5}
  \draw[grid] (\x,0)--(\x,5);
 \foreach \y in {.5,1.5,...,4.5}
  \draw[grid] (0,\y)--(11,\y);
 \foreach \xa/\xb in {.5/1.5,1.5/5.5,5.5/9.5,9.5/10.5}{
  \foreach \ya/\yb in {.5/2.5,2.5/4.5}
   \draw[map] (\xa,\ya) rectangle (\xb,\yb);
 }
 \draw[chosen,fill=none] (1.5,2.5) rectangle (5.5,4.5);
 \node[figgreen,fill=white,inner sep=2pt] at (3.5,3.5) {$I\times J$};
 \draw[line width=.8pt] (0,0) rectangle (11,5);
 \node[above=3pt] at (5.5,5) {$U\times V$};
 \draw[->,line width=.8pt] (0,-.25)--(11.6,-.25) node[right] {$U$};
 \draw[figgreen,line width=2pt] (1.5,-.25)--(5.5,-.25)
  node[midway,below=3pt] {$I$};
 \draw[->,line width=.8pt] (-.35,0)--(-.35,5.4) node[above] {$V$};
 \draw[figgreen,line width=2pt] (-.35,2.5)--(-.35,4.5)
  node[midway,left=3pt] {$J$};
 \node at (5.5,-1.14) {(b) A self-query from stored rectangles};
\end{scope}
\end{tikzpicture}
\caption{Preprocessing and self-queries. (a)~The index stores a map
for every product of canonical intervals. (b)~A query is partitioned
into canonical rectangles and thin end strips (orange); $I\times J$
is one stored rectangle.}
\label{fig:base-index}
\end{figure}
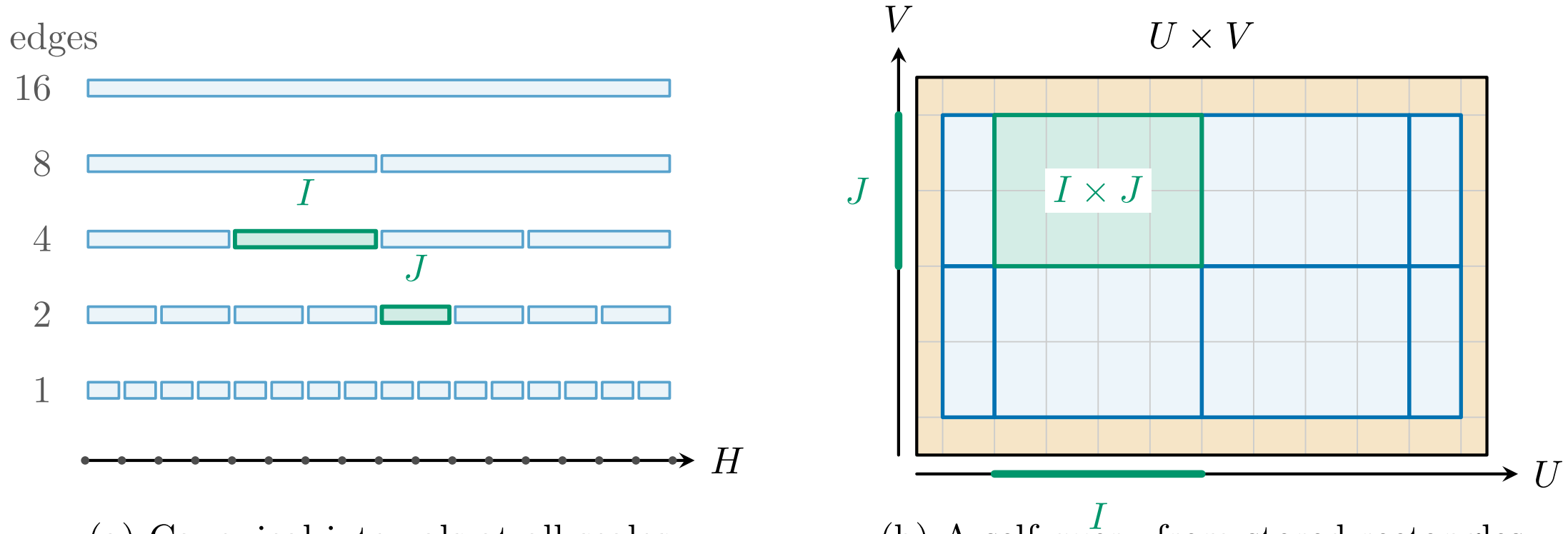

%% file: figures/shared-surrogate.tex
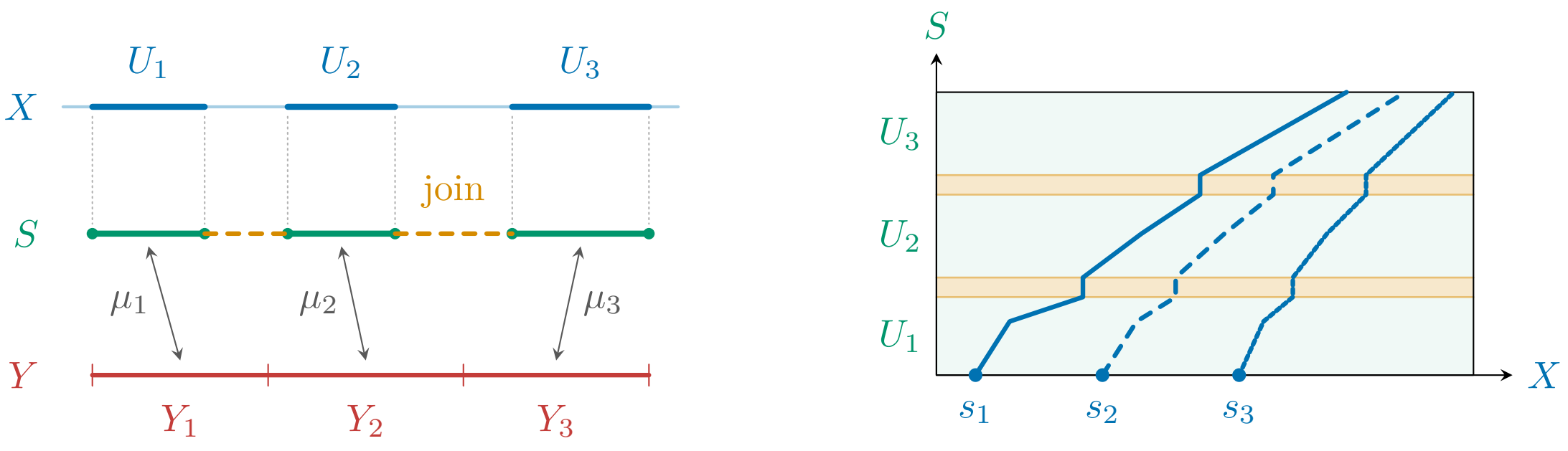
\begin{figure}[t]
\centering
\begin{tikzpicture}[x=0.9cm,y=0.9cm,>=stealth,line cap=round,line join=round,
  every node/.style={font=\small},
  selected/.style={line width=1.6pt},
  witness/.style={line width=1.2pt}]
\draw[figblue!35,line width=.8pt] (0,3.4)--(6.3,3.4);
\node[left,figblue] at (-.1,3.4) {$X$};
\node[left,figgreen] at (-.1,2.1) {$S$};
\node[left,figred] at (-.1,.65) {$Y$};
\foreach \a/\b/\i in {.3/1.45/1,2.3/3.4/2,4.6/6/3}{
  \draw[selected,figblue] (\a,3.4)--(\b,3.4)
    node[midway,above=3pt] {$U_{\i}$};
  \draw[selected,figgreen] (\a,2.1)--(\b,2.1);
  \draw[densely dotted,black!30] (\a,3.3)--(\a,2.2);
  \draw[densely dotted,black!30] (\b,3.3)--(\b,2.2);
  \fill[figgreen] (\a,2.1) circle (1.5pt);
  \fill[figgreen] (\b,2.1) circle (1.5pt);
}
\draw[dashed,witness,figorange] (1.45,2.1)--(2.3,2.1);
\draw[dashed,witness,figorange] (3.4,2.1)--(4.6,2.1);
\node[above=3pt,figorange] at (4,2.1) {join};
\draw[witness,figred] (.3,.65)--(6,.65);
\foreach \x in {.3,2.1,4.1,6}
  \draw[figred] (\x,.54)--(\x,.76);
\node[below=4pt,figred] at (1.2,.65) {$Y_1$};
\node[below=4pt,figred] at (3.1,.65) {$Y_2$};
\node[below=4pt,figred] at (5.05,.65) {$Y_3$};
\draw[<->,black!65] (.875,1.97)--(1.2,.8)
  node[midway,left] {$\mu_1$};
\draw[<->,black!65] (2.85,1.97)--(3.1,.8)
  node[midway,left] {$\mu_2$};
\draw[<->,black!65] (5.3,1.97)--(5.05,.8)
  node[midway,right] {$\mu_3$};
\node at (3,-.55) {(a) Donors and their matchings};
\begin{scope}[xshift=8.05cm]
\foreach \lo/\hi/\i in {.65/1.45/1,1.65/2.5/2,2.7/3.55/3}{
 \fill[figgreen!6] (0,\lo) rectangle (5.5,\hi);
 \node[left,figgreen] at (0,{(\lo+\hi)/2}) {$U_{\i}$};
}
\foreach \lo/\hi in {1.45/1.65,2.5/2.7}{
 \fill[figorange!20] (0,\lo) rectangle (5.5,\hi);
 \draw[figorange!55,line width=.5pt] (0,\lo)--(5.5,\lo);
 \draw[figorange!55,line width=.5pt] (0,\hi)--(5.5,\hi);
}
\draw (0,.65) rectangle (5.5,3.55);
\draw[->] (0,.65)--(5.9,.65) node[right,figblue] {$X$};
\draw[->] (0,.65)--(0,3.95) node[above,figgreen] {$S$};
\draw[witness,figblue] (.4,.65)--(.75,1.2)--(1.5,1.45)--(1.5,1.65)
 --(2.1,2.1)--(2.7,2.5)--(2.7,2.7)--(4.2,3.55);
\draw[witness,figblue,dashed] (1.7,.65)--(2.05,1.2)--(2.45,1.45)--(2.45,1.65)
 --(2.95,2.1)--(3.45,2.5)--(3.45,2.7)--(4.8,3.55);
\draw[witness,figblue,densely dotted] (3.1,.65)--(3.35,1.2)--(3.65,1.45)--(3.65,1.65)
 --(4.0,2.1)--(4.4,2.5)--(4.4,2.7)--(5.3,3.55);
\foreach \x/\i in {.4/1,1.7/2,3.1/3}{
  \fill[figblue] (\x,.65) circle (1.8pt);
  \node[below=3pt,figblue] at (\x,.65) {$s_{\i}$};
}
\node at (2.8,-.55) {(b) The same surrogate for all sources};
\end{scope}
\end{tikzpicture}
\caption{One surrogate serves many sources. (a)~The donor matchings $\mu_i$ and
the orange joins give a matching $\mu$ of $S$ and $Y$.
(b)~Each donor is queried against the whole of $X$. Paths from
different sources use the same rectangles $X\times U_j$ and cross
each orange join at a fixed position on $X$.}
\label{fig:shared-surrogate}
\end{figure}

%% file: sections/04-boundary-maps.tex

\section{Boundary maps}
\label{sec:boundary}

The recursions require two operations: constructing boundary
information for a rectangle and using it to propagate labels. We first
summarize reachability at $O(|R|)$ boundary points, then extend this
summary to a \emph{boundary map} on the entire boundary. Applying a
map takes $\tO(|R|)$ time, even when the input profiles have many
steps. We also give the operations needed to assemble profiles,
retain witnesses, and compose maps of adjacent rectangles. These tools
provide the boundary operations used in Sections~\ref{sec:query}
and~\ref{sec:index}.

Throughout this section, $R=[x_0,x_1]\times[y_0,y_1]$ is a rectangle
whose axes are subcurves of $P$ and $Q$ of sizes $p$ and $q$, and
$b=p+q=|R|$. Fix a radius $r\ge0$. Recall from
Section~\ref{sec:preliminaries} that the
original vertices subdivide each side of $R$ into $O(b)$ boundary
edges, and that the free part of each boundary edge is a free
interval, possibly empty. A point shared by two boundary
edges, such as a corner of $R$ or a vertex on a side, is treated as one
point for each incident edge, and we call the incident edge its
\emph{role}. On each side, we orient the parameter \emph{forward},
that is, upward on the left and right sides and rightward on the
bottom and top sides, so that forward motion along a side is monotone.

\subsection{Finite summaries}
\label{fd:finite}

A \emph{port} is a boundary point together with a role. We order the
input ports by the \emph{boundary order} $\preceq$ that starts at the
top-left corner, runs down the left side, and continues rightward
along the bottom side. Similarly, we order the output ports by the
order $\preceq$ that starts at the top-left corner, runs rightward
along the top side, and continues down the right side. Ports at the
same position are ordered by the order in which their incident edges
are encountered in the corresponding boundary traversal. Note that
$\preceq$ differs from the coordinatewise order $s\le t$, which means
$s_x\le_P t_x$ and $s_y\le_Q t_y$. The two boundary orders follow the
two arcs from the top-left to the bottom-right corner. Paths with
alternating endpoints therefore cross.

\begin{lemma}[Crossing]
\label{fd:crossing}
Let $s_1\preceq s_2$ be input points and $t_1\preceq t_2$ output
points of $R$, and let $\gamma_1$ and $\gamma_2$ be bi-monotone paths in
$R$ from $s_1$ to $t_2$ and from $s_2$ to $t_1$, respectively. Then
$\gamma_1$ and $\gamma_2$ intersect. Consequently, if both paths lie in
$\mathcal F_c$, then so do the bi-monotone paths from $s_1$ to $t_1$ and
from $s_2$ to $t_2$ obtained by splicing them at a common point. For
polygonal paths, a common point can be found in time linear in their
number of segments. For paths with steps in $\{(1,0),(0,1),(1,1)\}$
between grid points, it can be chosen to be a grid point.
\end{lemma}
\begin{proof}
The boundary orders and monotonicity give $s_i\le t_j$ for all
$i,j\in\{1,2\}$. Parameterize each path by $z=x+y$, omitting
stationary portions, and write $\gamma_k(z)=(x_k(z),z-x_k(z))$.
Their common parameter interval is $[z^-,z^+]$, where
\[
 z^-=\max_i(s_{i,x}+s_{i,y})
 \ \le\ z^+=\min_j(t_{j,x}+t_{j,y}).
\]
At $z^-$, either the first path is on the left side or the second is
on the bottom side, giving $x_1(z^-)\le_P x_2(z^-)$. At $z^+$,
either the first path is on the right side or the second is on the
top side, giving $x_1(z^+)\ge_P x_2(z^+)$. Continuity gives a common
point. Splicing there preserves monotonicity and membership in
$\mathcal F_c$.

For polygonal paths, merge the breakpoints in increasing $z$ and find
a zero of the piecewise-affine function $x_1-x_2$. This takes linear
time. Two intersecting grid steps from $\{(1,0),(0,1),(1,1)\}$ share
a grid point, proving the last claim.
\end{proof}

The crossing lemma explains why extremal entrances nearly determine
reachability. Suppose entrances $s_1$ and $s_3$ reach an exit $t$, and
$s_1\preceq s_2\preceq s_3$. If $s_2$ reaches any exit, its path
crosses one of the two paths to $t$, and a splice makes $t$ reachable
from $s_2$ as well (Figure~\ref{fig:boundary-crossing}). Thus we need
the extremal entrances for each exit and an \emph{activity} test for
each entrance. This use of extremal entrances is related to the
grid-reachability methods of Bringmann, K\"unnemann, and
Nusser~\cite{BKN}. We formalize the relation stored by our routines
as follows.

\input{figures/boundary-crossing}

\begin{definition}[Finite summary]
\label{def:summary}
Let $S_{\rm in}$ and $S_{\rm out}$ be finite sets of input and output
ports of $R$. A relation $B\subseteq S_{\rm in}\times S_{\rm out}$ is
an \emph{$(r,F)$-summary} if
\[
 \Rr r\cap(S_{\rm in}\times S_{\rm out})\ \subseteq\ B\ \subseteq\
 \Rr{Fr}\cap(S_{\rm in}\times S_{\rm out}),
\]
and if for every pair in $B$, a witness, that is, a bi-monotone path in
$R\cap\mathcal F_{Fr}$ between its ports, can be reported.
\end{definition}

Thus a finite summary is boundary information restricted to the
selected ports. To construct it, assign each input port its
\emph{rank} in $\preceq$ and run min- and max-propagation to find
extremal entrances. A third run uses equal labels on all output ports
and reverses both axes. This \emph{reverse activity} test marks every
input port that $r$-reaches an output port, and each marked port has
an $Fr$-witness to some output. The crossing lemma then fills in the
active entrances between the extremes.

\begin{lemma}[Compiling a finite summary]
\label{fd:completion}
Let $|S_{\rm in}|+|S_{\rm out}|=O(b)$, and suppose that
$(r,F)$-certified min-propagation and max-propagation of the ranks of
$S_{\rm in}$, and $(r,F)$-certified reverse activity from $S_{\rm
out}$, have been computed at the ports. Then an $(r,F)$-summary $B$ of
the form
\begin{equation}
 B(s,t)\quad\Longleftrightarrow\quad
 t\in T,\quad s\in H,\quad L(t)\le\rank(s)\le U(t),
 \label{fd:finiteformula}
\end{equation}
where $T\subseteq S_{\rm out}$, $H\subseteq S_{\rm in}$, and $L$ and
$U$ are nondecreasing on $T$ with respect to $\preceq$, can be computed
in $O(b)$ time. A witness for a pair of $B$ is obtained from at most
three witnesses of the three propagations by at most two splices, in
time linear in their sizes.
\end{lemma}
\begin{proof}
Let $T$ be the set of output ports that receive a finite label in the
min-propagation or in the max-propagation, and for $t\in T$, let $W(t)$
be the set of the one or two ranks received by $t$. Let $H$ be the set
of input ports that receive a label in the reverse-activity
propagation. We define
\[
 \ell(t)=\min W(t),\quad u(t)=\max W(t),\qquad
 L(t)=\min_{z\in T,\,z\succeq t}\ell(z),\qquad
 U(t)=\max_{z\in T,\,z\preceq t}u(z).
\]
A suffix-minimum sweep and a prefix-maximum sweep compute $L$ and $U$
in $O(b)$ time, together with ports attaining them. By definition, $L$
and $U$ are nondecreasing, and $L(t)\le\ell(t)\le u(t)\le U(t)$.

Let $z\succeq t$ attain $L(t)$, let $a$ have rank $L(t)$, and choose
$c$ with $\rank(c)\in W(t)$. Soundness gives witnesses $a\to z$ and
$c\to t$. Since $a\preceq c$ and $t\preceq z$, Lemma~\ref{fd:crossing}
gives an $Fr$-witness $a\to t$. The port of rank $U(t)$ reaches $t$
symmetrically.

Completeness of the three runs gives
\[
 s\Rr r t\quad\Longrightarrow\quad
 s\in H,\quad t\in T,\quad
 L(t)\le\ell(t)\le\rank(s)\le u(t)\le U(t),
\]
so $(s,t)\in B$. Conversely, for $(s,t)\in B$, activity supplies a
witness $s\to v$ for some output $v$. If $v\prec t$, splice it with
the witness from rank $L(t)$ to $t$. If $v\succ t$, use rank $U(t)$.
If $v=t$, no splice is needed. Each case uses at most three raw
witnesses and two applications of Lemma~\ref{fd:crossing}.
\end{proof}

Once this summary is stored, new labels can be propagated without
repeating the three runs. Each output requires only a range query on
the input ranks.

\begin{lemma}[Propagation at ports]
\label{fd:ports}
Given $B$ in the form~\eqref{fd:finiteformula} and labels on
$S_{\rm in}$, min- or max-propagation through $B$, with attaining
input ports for finite outputs, takes $O(b\log b)$ time. The propagation is
$(r,F)$-certified, with witnesses supplied by
Lemma~\ref{fd:completion}.
\end{lemma}
\begin{proof}
For min-propagation, set $a_{\rank(s)}=\lambda(s)$ for $s\in H$ and
$a_{\rank(s)}=+\infty$ otherwise. For $t\in T$, return
\[
 \lambda'(t)=\min_{L(t)\le j\le U(t)}a_j
\]
and an attaining port. Set $\lambda'(t)=+\infty$ for $t\notin T$.
A range-minimum tree answers the $O(b)$ queries in $O(b\log b)$ time.
Certification follows from $\Rr r\subseteq B\subseteq\Rr{Fr}$ on the
ports. Max-propagation is symmetric.
\end{proof}

\subsection{From ports to continuous boundaries}
\label{fd:normalization}

A continuous boundary has infinitely many points, but only $O(b)$ free
intervals, so we let one port represent each free interval. For every
input boundary edge $e$ with nonempty free interval $I_e$, let $p_e$ be
the first point of $I_e$ in the forward direction, and for every
output boundary edge $g$ with nonempty free interval $I_g$, let $z_g$
be the last point of $I_g$. For a relation $B$ on these
\emph{representative ports}, we define the relation $\Lift(B)$ on free
boundary points by
\begin{equation}
 \Lift(B)(s,t)\quad\Longleftrightarrow\quad
 s\in I_e,\ t\in I_g,\ s\le t,\ \text{and}\ B(p_e,z_g)
 \ \text{for some edges $e$ and $g$.}
 \label{fd:liftformula}
\end{equation}
Choosing the first point of an input interval and the last point of an
output interval makes every true crossing extend to the
representatives. In the other direction, a path between representatives
must be adjusted to the actual points, and this costs at most $2r$ on
each axis.

\begin{lemma}[Lift]
\label{fd:lift}
If $B$ is an $(r,F)$-summary on the representative ports, then
$\Rr r\subseteq\Lift(B)\subseteq\Rr{(F+4)r}$ on pairs of free input
and output points. More precisely, if $s\in I_e$, $t\in I_g$, and
$s\le t$, then clamping a witness $\gamma$ for $(p_e,z_g)$ coordinatewise
to the box $[s_x,t_x]\times[s_y,t_y]$ yields a witness for $(s,t)$ of
cost at most $(F+4)r$, in time linear in the size of $\gamma$.
\end{lemma}
\begin{proof}
If $s\Rr r t$, extend an $r$-witness by the free boundary portions
$p_e\to s$ and $t\to z_g$. Then $p_e\Rr r z_g$, so $B(p_e,z_g)$ and
$\Lift(B)(s,t)$ hold.

Conversely, coordinatewise clamping preserves continuity and
monotonicity and sends the endpoints of $\gamma$ to $s$ and $t$.
For a point $(x,y)$ of $\gamma$, let $(x',y')$ be its clamp. If
$x<_P s_x$, then $e$ is a bottom edge and $x,x'\in I_e$. If
$x>_P t_x$, then $g$ is a top edge and $x,x'\in I_g$. In either case,
$P(x)$ and $P(x')$ lie within $r$ of the same point of $Q$, giving
$\|P(x)-P(x')\|\le2r$. Symmetrically,
$\|Q(y)-Q(y')\|\le2r$. Thus
\[
 \|P(x')-Q(y')\|\le\|P(x')-P(x)\|+\|P(x)-Q(y)\|+\|Q(y)-Q(y')\|
 \le(F+4)r.
\]
Clamping splits each segment of $\gamma$ into $O(1)$ segments, which
gives the time bound.
\end{proof}

We also allow free motion along the input boundary before a crossing
and along the output boundary afterwards. Besides preserving
reachability, these motions make the label profiles monotone. Let
$C_{\rm in}$ be the relation of pairs $(s,s')$ of input points such
that some bi-monotone path from $s$ to $s'$ lies in $\In(R)\cap\mathcal
F_r$, and define $C_{\rm out}$ analogously with $\Out(R)$. Thus
$C_{\rm in}$ moves forward along one input side and may pass from the
bottom-left corner onto either side, and $C_{\rm out}$ moves forward
along one output side and may enter the top-right corner from either
side.

\begin{definition}[Boundary map]
\label{def:boundary-map}
For an $(r,F)$-summary $B$ on the representative ports of $R$, the
\emph{boundary map} induced by $B$ is
$T_B=C_{\rm out}\circ\Lift(B)\circ C_{\rm in}$. We call $F$ its
\emph{raw factor} and $F+4$ its \emph{factor}. Min- and
max-propagation through $T_B$ take, respectively, the minimum and
maximum of $\lambda(s)$ over $(s,t)\in T_B$, with the unavailable
labels of Section~\ref{sec:preliminaries} for empty sets.
\end{definition}

Witness reporting requires a record of these three parts. A pair
$(s,t)\in T_B$ has a
\emph{decomposition} $(s,s',t',t,e,g)$, where $(s,s')\in C_{\rm in}$,
$s'\in I_e$, $t'\in I_g$, $s'\le t'$, $B(p_e,z_g)$, and
$(t',t)\in C_{\rm out}$. Every record created when applying a map
stores such a decomposition. The next lemma applies to all retained
records, including those that do not attain the selected extremum.

\begin{lemma}[Boundary maps]
\label{fd:closure}
\label{fd:rawreports}
Let $B$ be an $(r,F)$-summary on the representative ports of $R$. Then
$\Rr r\subseteq T_B\subseteq\Rr{(F+4)r}$. Hence $T_B$ is
$(r,F+4)$-boundary information of $R$, and both min- and max-propagation
through $T_B$ are $(r,F+4)$-certified. Given a decomposition
$(s,s',t',t,e,g)$ of a pair
of $T_B$ and a witness $\gamma$ for $(p_e,z_g)$, a witness for $(s,t)$
of cost at most $(F+4)r$ is obtained in time linear in the size of
$\gamma$ plus the number of boundary edges between $s$ and $s'$ and
between $t'$ and $t$.
\end{lemma}
\begin{proof}
Both closures contain the identity on free points, so
$\Rr r\subseteq\Lift(B)\subseteq T_B$ by Lemma~\ref{fd:lift}. For the
second inclusion, concatenate the boundary path $s\to s'$, the clamp
of $\gamma$ to $[s'_x,t'_x]\times[s'_y,t'_y]$, and the boundary path
$t'\to t$. Their costs are at most $r$, $(F+4)r$, and $r$, respectively,
so the concatenation has cost at most $(F+4)r$. Lemma~\ref{fd:lift}
and a scan of the boundary portions give the reporting bound.
\end{proof}

Each map adds $4r$ to the cost of its raw summary witnesses. The raw
factor $F$ already includes any losses incurred in constructing that
summary. A record points to a raw witness and applies this map's
clamp once, at the recorded anchors $s'$ and $t'$.

\subsection{Applying a boundary map to label profiles}
\label{fd:application}

It remains to apply a boundary map efficiently. Recall that labels
consist of a weight and a source identifier, ordered lexicographically,
and that a profile is a step function on a boundary edge. Profiles
may accumulate many steps as queries are composed. We avoid copying
these steps by sharing unchanged portions of inherited profiles.
Forward closure gives the monotonicity needed for this representation.

\begin{definition}[Monotone profiles]
\label{def:monotone-profile}
A \emph{min-profile} on a boundary edge $e$ is a function from $I_e$
to labels and $+\infty$ that is nonincreasing in the forward direction,
piecewise constant with finitely many pieces, and whose finite part is
a possibly empty closed final segment of $I_e$. A \emph{max-profile} is defined in
the same way, with a nondecreasing function and unavailable value
$-\infty$.
\end{definition}

Profiles retain the source and witness records associated with their
finite values. An output record created by applying a boundary map
stores a decomposition as above. Inherited records are shared.

Under forward closure, the set of sources reaching a point grows as
the point moves forward, so the minimum label is nonincreasing. Below
we show how to compute this closure for profiles and for labels given
on closed intervals or points.

Fix a top output edge $g$ with $I_g\ne\varnothing$ in the original
edge $[i,i+1]$ of $P$, and let $\varphi_e$ denote the min-profile of an
input edge $e$ after input closure. Let $E_g$ be the set of input
edges with nonempty free intervals on the left side or on the bottom
portion $[x_0,i]$. Let $e^*$ be the bottom edge in $[i,i+1]$.
Write $I_{e^*}=[a^*,b^*]$ when it is nonempty. Define
\begin{equation}
 \begin{aligned}
 c_g&=\min\{\varphi_e(\max I_e): e\in E_g,\ B(p_e,z_g)\},\\
 h_g(x)&=
 \begin{cases}
 \varphi_{e^*}(\min\{x,b^*\}), & x\ge_P a^*\text{ and }B(p_{e^*},z_g),\\
 +\infty, & \text{otherwise,}
 \end{cases}
 \end{aligned}
 \label{fd:onefunction}
\end{equation}
and set $h_g\equiv+\infty$ if $I_{e^*}$ is empty. All eligible edges
except $e^*$ contribute a scalar minimum, while $e^*$ contributes
one inherited profile. Right output edges are treated symmetrically.

\begin{lemma}[One inherited function]
\label{fd:formula}
For every top output edge $g$ with $I_g\ne\varnothing$ and every
$(x,y_1)\in I_g$ with $x<_P i+1$,
the min-propagation of the closed input profiles through $\Lift(B)$,
restricted to pairs with role $g$, equals $\min\{c_g,h_g(x)\}$. At the
$O(b)$ output points at original vertices, the propagation equals the
minimum of this expression and $O(1)$ further terms of the same kind.
Given the endpoint values of the input profiles, all $c_g$ and vertex
terms can be computed in $O(b\log b)$ time from
\eqref{fd:finiteformula}.
\end{lemma}
\begin{proof}
Let $t=(x,y_1)$ with $x<_P i+1$. For $e\in E_g$ with $B(p_e,z_g)$,
every $s\in I_e$ satisfies $s\le t$, and monotonicity gives
$\min_{s\in I_e}\varphi_e(s)=\varphi_e(\max I_e)$. For a nonempty
$I_{e^*}$, the eligible interval is $[a^*,\min\{x,b^*\}]$, giving
$h_g(x)$, with value $+\infty$ when this interval is empty.
No other bottom edge contributes. At $x=i+1$, the next bottom edge
and the other roles at that vertex add $O(1)$ terms.

The ranks of $E_g$ form a prefix of the input order. Thus $c_g$ is a
range minimum on $[L(z_g),U(z_g)]$ intersected with that prefix,
with ports outside $H$ assigned $+\infty$. A range-minimum tree on
the terminal values answers all $O(b)$ queries, with attaining
edges, in $O(b\log b)$ time. The vertex terms use the given endpoint
values. For right output edges, the corresponding ranks form a suffix.
\end{proof}

Only the bottom edge directly below a top output contributes a
varying profile. The other eligible edges contribute scalar minima.
Combining a nonincreasing profile with a scalar changes only a prefix.
For example, the steps $9,6,2$ become $7,6,2$ when combined with
$c=7$, and the suffix $6,2$ can be shared
(Figure~\ref{fig:continuous-profile}).

\input{figures/continuous-profile}

We store each profile in a persistent balanced search
tree~\cite{DSST89} keyed by breakpoints, in which every node stores the
extreme coordinates and labels of its subtree. Separately, each
generated record stores its label, the operation that created it, and
predecessor pointers with the attained points used by that operation.
Let $M$ bound the total number of records and tree nodes allocated by
the algorithm. The numerical trees share
unchanged parts of step functions, while the separate records keep
track of where each contribution came from.

These records implement the local witnesses of
Definition~\ref{def:certified}. Dependencies are created in
topological order, and each primitive operation can report its path.
Traceback follows predecessor pointers, preserving the label and
joining paths at the stored points, until it reaches an input of the
current invocation. It takes time proportional to the total reporting
time of the visited operations and does not expand the earlier history
of the input label.

\begin{lemma}[Persistent profile operations]
\label{fd:tree}
On a min-profile, the following operations take $O(\log M)$ time and
new nodes: evaluation, finding the first or last finite point,
restriction to an interval, concatenation on adjacent intervals,
constant extension, and taking the minimum with a constant on a
final segment. Concatenation and extension must preserve monotonicity.
The bounds are independent of the number of steps in the profile.
The same holds for max-profiles.
\end{lemma}
\begin{proof}
Searching, splitting, and joining a persistent balanced tree copy only
$O(\log M)$ nodes and share all other subtrees. Evaluation and finite
extrema use one search, restriction uses two splits, and concatenation
and extension use a join. For the minimum with $c$ on a final segment,
split at its start. Monotonicity makes the set where $c$ is smaller a
prefix. Locate its end by a search in label order, replace the prefix
by one constant piece, and share the suffix.
\end{proof}

Restrictions may leave gaps in a profile. Before using such information
on another boundary, we close it under forward motion through the free
intervals. The same operation inserts the scalar records produced when
we transfer outputs between curves.

\begin{lemma}[Boundary assembly]
\label{fd:boundary-assembly}
Let $\Gamma\in\{\In(R),\Out(R)\}$, with corresponding closure
$C_\Gamma\in\{C_{\rm in},C_{\rm out}\}$. On each boundary edge, suppose we
are given at most one monotone profile restricted to a closed
subinterval of its free interval, together with $L$ scalar-labelled
closed intervals and points in total, each contained in one free
boundary interval. Let $\eta$ be the pointwise
minimum of these contributions, with value $+\infty$ elsewhere. Then
its forward closure
\[
 \bar\eta(t)=\min\{\eta(s):(s,t)\in C_\Gamma\}
\]
can be computed as min-profiles in $O((b+L)\log M)$ time and new
nodes. Each finite output retains a supplied record and an $r$-free
boundary path from a point in its support. Extending a supplied witness
of cost $c$ therefore gives a witness of cost at most $\max\{c,r\}$.
The analogous statement holds for max-profiles.
\end{lemma}
\begin{proof}
Consider a free interval $I=[a,b']$ and an inherited profile $\phi$
restricted to $J=[u,v]\subseteq I$. Its closure within $I$ is
\[
 \widehat\phi(x)=
 \begin{cases}
  +\infty,&x\in[a,u),\\
  \phi(\min\{x,v\}),&x\in[u,b'].
 \end{cases}
\]
Restriction and constant extension construct this profile in
$O(\log M)$ time by Lemma~\ref{fd:tree}. A scalar record of label
$\ell$ on $[u,v]$ contributes $\ell$ on $[u,b']$, so each scalar
record requires one minimum with a constant on a final segment.

Sweep each boundary side forward, carrying the last label of an edge
to the next edge exactly when their common vertex is free. At a corner
shared by the two sides, combine its incident values and propagate
them in the permitted forward directions. There are $O(b+L)$ profile
operations. Each operation stores its attaining record and boundary
predecessor, and all added boundary portions lie in $\mathcal F_r$.
\end{proof}

For fixed recursion depths, our algorithms perform polynomially many
profile operations, so $M$ is polynomial in $m+n$ and $\log M$ is
absorbed in $\tO(\cdot)$. We can now combine the inherited-function
formula with boundary assembly to apply a map in time proportional,
up to logarithmic factors, to its boundary size.

\begin{lemma}[Applying a boundary map]
\label{fd:profiles}
Let $B$ be an $(r,F)$-summary on the representative ports of $R$, given
in the form~\eqref{fd:finiteformula}. Given min-profiles on the input
edges, the min-propagation through $T_B$ can be computed as min-profiles
on the output edges in $O(b\log(b+M))$ time. The same bound holds for
max-propagation of max-profiles. Every created record stores a
decomposition, so it has a local witness of cost at most $(F+4)r$ by
Lemma~\ref{fd:rawreports}.
\end{lemma}
\begin{proof}
Apply $C_{\rm in}$ by Lemma~\ref{fd:boundary-assembly}. Compute the
endpoint values, then $c_g$ and the vertex terms for every output edge
using Lemma~\ref{fd:formula}. Restriction and constant extension form
$h_g$, and one minimum with a constant gives $\min\{c_g,h_g\}$.
Finally apply $C_{\rm out}$ by Lemma~\ref{fd:boundary-assembly}.
The range queries take $O(b\log b)$ time, and the $O(b)$ profile
operations take $O(b\log M)$ time.

Range queries retain attaining ports, inherited functions retain the
coordinate rule $x\mapsto\min\{x,b^*\}$, and closures retain boundary
predecessors. Together these specify each record's decomposition.
Max-propagation is symmetric.
\end{proof}

For labels entering through only one side, outputs on the opposite
side have an especially short description. This is the form used by
the inner recursion when transferring outputs to a different curve.

\begin{lemma}[Opposite-side records]
\label{fd:opposite-records}
Suppose that a boundary map is applied to labels finite only on the
bottom side of $R$. On each right output edge, its min-propagation is
the minimum of $O(1)$ scalar records. Each record consists of a label,
a closed interval on which it is valid, and a decomposition of a pair
of the map for every point of that interval. The $O(b)$ records can be
retained while applying the map in $O(b\log(b+M))$ time, including
records that do not attain the minimum. Every retained record has the
witness guarantee of Lemma~\ref{fd:rawreports}. The statement also
holds for max-propagation and with the axes exchanged.
\end{lemma}
\begin{proof}
Every bottom point precedes every right output coordinatewise. After
$C_{\rm in}$, the only finite labels on the left side are copies of
the bottom-left corner label, so each left-edge profile is constant
on its finite support. In Lemma~\ref{fd:formula} with the axes
exchanged, $c_g$ is therefore a scalar record and $h_g$ is constant on
its support. Output closure adds at most one scalar carried from the
preceding right edge. Vertex roles and the value carried from the top
side to the top-right corner add $O(1)$ singleton records. All
supports are closed intervals. Retaining these contributions and their
decompositions adds $O(1)$ records per edge to the computation in
Lemma~\ref{fd:profiles}.
\end{proof}

The same machinery turns a certified propagation routine into a
reusable boundary map. Only three runs with fresh labels are needed.
We retain their states so that constructing the map does not require
reporting its witnesses.

\begin{lemma}[Compiling a routine]
\label{ix:compiler}
Let $\mathcal A$ compute $(r,F)$-certified min- and max-propagation on
$R$ and its reversal, for labels at $O(b)$ ports, with search choices
and boundary relations independent of those labels. If its running
time is $\tau_{\rm run}\ge b$ and its local witness-reporting time is
$\tau_{\rm wit}$, then an $(r,F)$-summary $B$ of size $O(b)$ can be
constructed in $O(\tau_{\rm run})$ time, retaining the required
witness states. The resulting map $T_B$ has raw factor $F$ and factor
$F+4$. Applying it to input profiles takes $O(b\log(b+M))$ time, and
a witness for any retained record can be reported in
$O(\tau_{\rm wit}+b)$ time.
\end{lemma}
\begin{proof}
Run $\mathcal A$ for min- and max-propagation of input-port ranks and
for reverse activity from output ports. Lemma~\ref{fd:completion}
constructs $B$ from these runs in
$3\tau_{\rm run}+O(b)=O(\tau_{\rm run})$ time.
Lemma~\ref{fd:profiles} gives the application bound. By
Lemmas~\ref{fd:completion} and~\ref{fd:rawreports}, each retained
record is witnessed by at most three reported paths, at most two
splices, one clamp, and boundary paths of total size $O(b)$. Thus
its report takes $O(\tau_{\rm wit}+b)$ time and has cost at most
$(F+4)r$.
\end{proof}

\subsection{Networks of rectangles}
\label{ix:dag-section}

We next combine the information stored for adjacent rectangles.
The witnesses concatenate, so crossing more rectangles does not
increase the approximation factor. We first state this property for
boundary relations, then implement it by forwarding label profiles.

\begin{lemma}[Boundary information composes]
\label{pre:compose}
Let $R=R_1\cup\cdots\cup R_N$ for rectangles $R_i$ with disjoint
interiors, and let $\mathcal B_i$ be $(r,C)$-boundary information of
$R_i$. Then
\[
 \mathcal B=(\In(R)\times\Out(R))
 \cap(\mathcal B_1\cup\cdots\cup\mathcal B_N)^+
\]
is $(r,C)$-boundary information of $R$, where the superscript $+$
denotes transitive closure. A witness is obtained by concatenating the
local witnesses along a chain in this closure.
\end{lemma}
\begin{proof}
Let $\pi:s\to t$ lie in $R\cap\mathcal F_r$. For positive-area $R$,
discard zero-area pieces and write
$R_i=[x_0^i,x_1^i]\times[y_0^i,y_1^i]$. Assign each point to the unique
half-open box $[x_0^i,x_1^i)\times[y_0^i,y_1^i)$,
closing a side when it lies on $\Out(R)$. Once $\pi$ leaves a box,
it keeps $x\ge_P x_1^i$ or $y\ge_Q y_1^i$ and cannot return.
Thus $\pi$ splits into portions
$\pi_j:u_{j-1}\to u_j$ inside $R_{i_j}$, with $\ell\le N$,
$u_0=s$, and $u_\ell=t$. Each portion enters through
$\In(R_{i_j})$ and leaves through $\Out(R_{i_j})$, so
\[
 (u_{j-1},u_j)\in\mathcal B_{i_j}\quad(1\le j\le\ell)
 \quad\Longrightarrow\quad(s,t)\in\mathcal B.
\]
For a segment $R$, subdivide at the piece endpoints and use
$\In(R_i)=\Out(R_i)=R_i$. A point requires no subdivision.
This proves $\Rr r\subseteq\mathcal B$.

Conversely, concatenate the witnesses of any chain defining a pair
of $\mathcal B$. The resulting path is bi-monotone and has cost
$\max_j\operatorname{cost}(\pi_j)\le Cr$, proving
$\mathcal B\subseteq\Rr{Cr}$.
\end{proof}

For the implementation, suppose $R$ is partitioned into positive-area
rectangles $R_1,\ldots,R_N$. Number them so that $i<j$ whenever an
output side of $R_i$ and an input side of $R_j$ share a segment of
positive length. A grid is ordered by the sum of its row and column
indices. Each rectangle has an $(r,C)$-certified propagation routine.

\emph{Network propagation} processes the rectangles in this order.
For min-propagation, the input at a point is the minimum of the
original input label, when present, and the labels returned there by
earlier rectangles. Write $\lambda_i$ for this labelling on $\In(R_i)$
after forward closure, and $\eta_i$ for the local output. The final
answer is the minimum of the $\eta_i$ on $\Out(R)$, again followed by
forward closure. Max-propagation uses maxima throughout.

\begin{lemma}[Networks]
\label{ix:dag}
Network propagation is $(r,C)$-certified on $R$. A reported witness
concatenates local witnesses and free boundary portions, using each
local routine at most once. For a $u\times v$ grid, it uses
$O(u+v)$ local witnesses. If the local routines accept and return
monotone profiles in times $\tau_i$, the total propagation time is
\[
 O\!\left(\sum_{i=1}^N\tau_i+S\log(S+M)\right),
 \qquad S=\sum_{i=1}^N|R_i|.
\]
\end{lemma}
\begin{proof}
We prove min-propagation. For an $r$-path $\pi:s\to t$, use the
half-open decomposition from Lemma~\ref{pre:compose}, with portions
$\pi_j:u_{j-1}\to u_j$ in $R_{i_j}$. The indices $i_j$ increase.
This follows from the numbering when consecutive pieces share an
edge. If they meet only at a corner, the rectangle immediately
southeast of that corner has an index strictly between theirs.
Completeness of the local routines therefore gives, inductively,
\[
 \eta_{i_j}(u_j)\le\lambda_{i_j}(u_{j-1})\le\lambda(s).
\]
Forward closure and the final minimum preserve this inequality.

For a finite output, traceback preserves the label and concatenates
local witnesses with $r$-free boundary paths. Their costs are at most
$Cr$, so the concatenation has cost at most $Cr$. Call indices
decrease during traceback, and hence no local routine is used twice.
In a grid, subdividing a bi-monotone path at the grid lines gives
$O(u+v)$ pieces. Each piece is incident to $O(1)$ cells, including
edge and vertex contacts, which bounds the number of local witnesses.

For the time bound, refine boundary edges at the $O(N)$ partition
junctions, obtaining $O(S)$ pieces. On each open input piece there is
one inherited profile, from a predecessor or from the global input.
Only junctions require scalar minima. Restrict these profiles and
sweep each input side forward, carrying an endpoint label across a
free shared endpoint. Each piece uses $O(1)$ restrictions, scalar
minima, and concatenations from Lemma~\ref{fd:tree}. The final output
is assembled in the same way. Sorting and processing the pieces costs
\[
 O(S\log S)+O(S\log M)=O(S\log(S+M)),
\]
in addition to the local calls. Max-propagation is symmetric.
\end{proof}

\subsection{The exact base}
\label{fd:exactbase}

The recursions start from exact propagation through the cells of the
free-space diagram. Convexity gives the local transfer, and the
network lemma combines these transfers without approximation loss.
This extends the algorithms of Alt and Godau~\cite{AG95} and Eiter
and Mannila~\cite{EM94} to the label interface used here.

\begin{lemma}[Exact propagation]
\label{fd:cell}
Let $R$ have subcurve sizes $p,q$. Exact, that is,
$(r,1)$-certified, min- or max-propagation through $R$ takes
$O(pq\log M)$ time, and a local
witness with $O(p+q)$ segments can be reported in $O((p+q)\log M)$
time. In the discrete setting, the time bounds are $O(pq)$ and
$O(p+q)$.
\end{lemma}
\begin{proof}
We describe min-propagation. In a possibly clipped cell, convexity gives
\[
 s\Rr r t\quad\Longleftrightarrow\quad
 s,t\in\mathcal F_r\text{ and }s\le t.
\]
For a top output, all free left inputs are eligible, and the eligible
bottom inputs form a prefix. Its label is therefore the minimum of a
scalar and one restricted profile, with constant extension where
needed. The right side is symmetric. Each cell uses $O(1)$ operations
of Lemma~\ref{fd:tree}, with straight-segment witnesses.
Lemma~\ref{ix:dag} with $C=1$ gives $O(pq\log M)$ propagation time
and $O((p+q)\log M)$ reporting time through the cell grid.
For a segment or point $R$, a forward scan takes $O((p+q)\log M)$
time, within the same bound since $\min\{p,q\}\le2$.

In the discrete setting, set unavailable and missing labels to
$+\infty$. At a free state, the recurrence is
\[
 L(i,j)=\min\{\lambda(i,j),L(i-1,j),L(i,j-1),L(i-1,j-1)\},
\]
with $\lambda(i,j)=+\infty$ off the input boundary and $L(i,j)=+\infty$
at nonfree states. Induction on $i+j$ proves exactness. There are
$O(pq)$ states and $O(p+q)$ traceback steps. Max-propagation uses
maxima and unavailable value $-\infty$.
\end{proof}

%% file: figures/boundary-crossing.tex
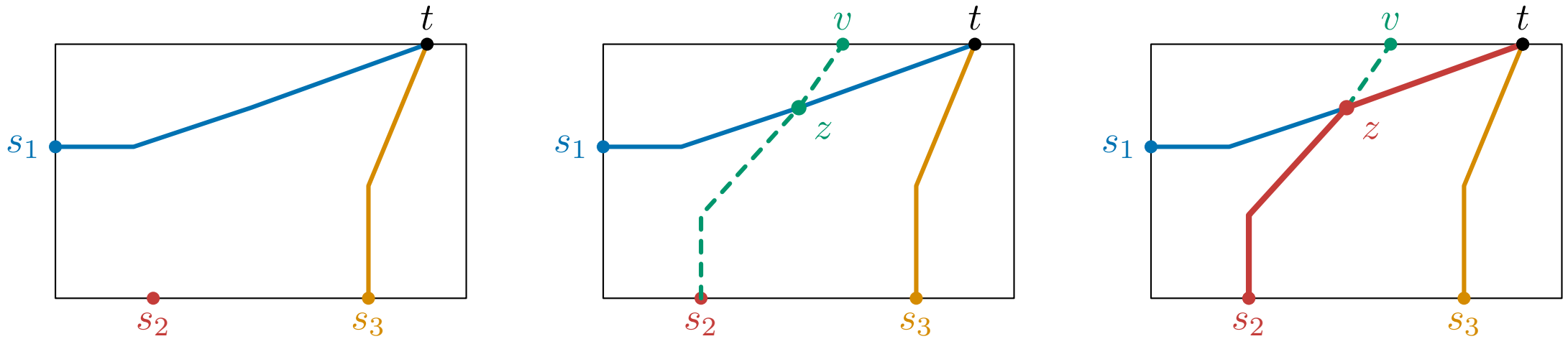
\begin{figure}[t]
\centering
\begin{tikzpicture}[x=0.9cm,y=0.9cm,>=stealth,line cap=round,line join=round,
  every node/.style={font=\small},
  witness/.style={line width=1.15pt}]
\foreach \shift in {0,5.6,11.2}{
 \begin{scope}[shift={(\shift,0)}]
 \draw (0,0) rectangle (4.2,2.6);
 \draw[witness,figblue] (0,1.55)--(.8,1.55)--(2,1.95)--(3.8,2.6);
 \draw[witness,figorange] (3.2,0)--(3.2,1.15)--(3.8,2.6);
 \fill[figblue] (0,1.55) circle (1.7pt) node[left] {$s_1$};
 \fill[figred] (1,0) circle (1.7pt) node[below] {$s_2$};
 \fill[figorange] (3.2,0) circle (1.7pt) node[below] {$s_3$};
 \fill (3.8,2.6) circle (1.7pt) node[above] {$t$};
 \end{scope}
}
\node at (2.1,-.8) {(a) Extremal entrances};
\begin{scope}[shift={(5.6,0)}]
 \draw[witness,dashed,figgreen] (1,0)--(1,.85)--(2,1.95)--(2.45,2.6);
 \fill[figgreen] (2.45,2.6) circle (1.7pt) node[above] {$v$};
 \fill[figgreen] (2,1.95) circle (2pt) node[below right] {$z$};
 \node at (2.1,-.8) {(b) An active entrance};
\end{scope}
\begin{scope}[shift={(11.2,0)}]
 \draw[witness,dashed,figgreen] (2,1.95)--(2.45,2.6);
 \draw[line width=1.55pt,figred] (1,0)--(1,.85)--(2,1.95)--(3.8,2.6);
 \fill[figgreen] (2.45,2.6) circle (1.7pt) node[above] {$v$};
 \fill[figred] (1,0) circle (1.7pt);
 \fill[figred] (2,1.95) circle (2pt) node[below right] {$z$};
 \fill (3.8,2.6) circle (1.7pt);
 \node at (2.1,-.8) {(c) Splicing at $z$};
\end{scope}
\end{tikzpicture}
\caption{Extremal entrances and activity. (a)~The two paths to $t$
alone give no path from $s_2$. (b)~A path from $s_2$ to an exit $v$
crosses an extremal path at $z$. (c)~The red splice reaches $t$ with
no larger radius. Here $v$ precedes $t$ in output order; an exit after
$t$ uses the other extremal path.}
\label{fig:boundary-crossing}
\end{figure}

%% file: figures/continuous-profile.tex
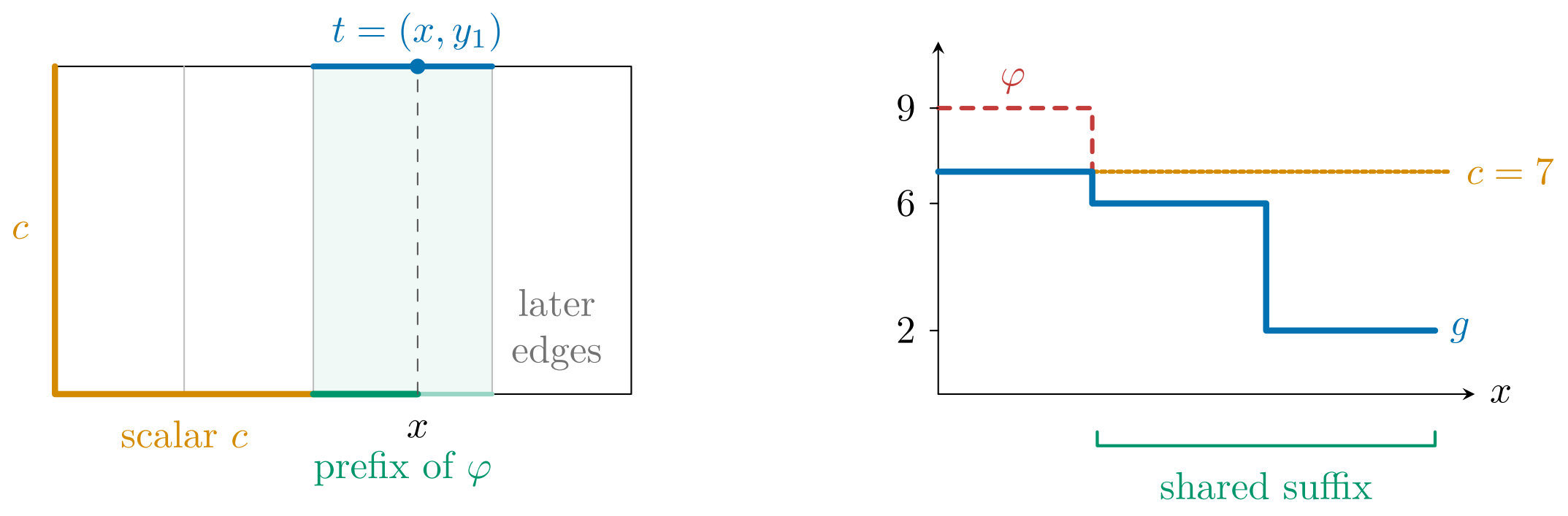
\begin{figure}[t]
\centering
\begin{tikzpicture}[x=0.9cm,y=0.9cm,>=stealth,line cap=round,line join=round,
  every node/.style={font=\small}]
\fill[figgreen!6] (2.6,.4) rectangle (4.4,3.7);
\draw (0,.4) rectangle (5.8,3.7);
\foreach \x in {1.3,2.6,4.4}
  \draw[black!25] (\x,.4)--(\x,3.7);
\draw[line width=1.6pt,figorange] (0,3.7)--(0,.4)--(2.6,.4);
\draw[line width=1.2pt,figgreen!40] (2.6,.4)--(4.4,.4);
\draw[line width=1.7pt,figgreen] (2.6,.4)--(3.65,.4);
\draw[line width=1.5pt,figblue] (2.6,3.7)--(4.4,3.7);
\draw[dashed,black!65] (3.65,.4)--(3.65,3.7);
\fill[figblue] (3.65,3.7) circle (2pt) node[above] {$t=(x,y_1)$};
\node[below=3pt] at (3.65,.4) {$x$};
\node[below=3pt,figorange] at (1.3,.4) {scalar $c$};
\node[figgreen] at (3.5,-.35) {prefix of $\varphi$};
\node[black!55,align=center] at (5.05,1.05) {later\\edges};
\node[left=3pt,figorange] at (0,2.05) {$c$};
\node at (2.9,-1.2) {(a) One varying input profile};
\begin{scope}[xshift=8cm]
\draw[->] (0,.4)--(5.4,.4) node[right] {$x$};
\draw[->] (0,.4)--(0,3.95);
\draw[densely dotted,line width=1.1pt,figorange] (0,2.64)--(5.15,2.64)
  node[right] {$c=7$};
\draw[dashed,line width=1.15pt,figred] (0,3.28)--(1.55,3.28)--(1.55,2.32)
 --(3.3,2.32)--(3.3,1.04)--(5,1.04);
\node[above,figred] at (.75,3.28) {$\varphi$};
\draw[line width=1.6pt,figblue] (0,2.64)--(1.55,2.64)--(1.55,2.32)
 --(3.3,2.32)--(3.3,1.04)--(5,1.04);
\node[right,figblue] at (5,1.04) {$g$};
\foreach \y/\v in {3.28/9,2.32/6,1.04/2}{
 \draw (0,\y)--(-.08,\y);
 \node[left=2pt] at (0,\y) {$\v$};
}
\draw[figgreen,line width=.8pt] (1.6,.02)--(1.6,-.12)--(5,-.12)--(5,.02);
\node[below=3pt,figgreen] at (3.3,-.12) {shared suffix};
\node at (2.65,-1.2) {(b) $g=\min\{7,\varphi\}$};
\end{scope}
\end{tikzpicture}
\caption{Updating a continuous label profile. (a)~Among the related
input edges, only the edge below $t$ contributes a varying prefix;
earlier edges contribute scalar minima. (b)~Taking the minimum with
$c=7$ changes $9,6,2$ to $7,6,2$. The unchanged suffix is shared with
$\varphi$.}
\label{fig:continuous-profile}
\end{figure}

%% file: sections/05-query-recursion.tex

\section{The inner recursion}
\label{sec:query}

The inner recursion turns a self-query routine into a general query
between separately indexed curves. We prove the guarantees used in
Section~\ref{intro:surrogates}: the cursor search constructs one
surrogate for all sources, self-queries propagate their labels through
it, and the surrogate matching carries the outputs back to the original
rectangle. The boundary operations are supplied by
Section~\ref{sec:boundary}. Throughout this section, the outer round
$i$ is fixed and omitted from the notation:
$\Query_t$, $\beta_t$, and $A_t$ denote $\Query_{i,t}$,
$\beta_{i,t}$, and $A_{i,t}$.

\subsection{The interface and query guarantee}
\label{sec:currency}

We use the profiles and witness records of
Section~\ref{fd:application}. Profiles are passed by reference.
Restrictions and newly supplied labels are closed forward along the
free boundary using Lemma~\ref{fd:boundary-assembly}. A local witness
starts at an input point of the current invocation, without expanding
the earlier history of its label.

An index packages boundary maps for one host curve. Its type records
the excess exponents of preprocessing and self-query time, together
with its approximation factor.

\begin{definition}[Index]
\label{def:index}
Let $0\le\delta\le\alpha\le1$ and $D\ge5$. An \emph{index of type
$(\alpha,\delta;D)$} provides two deterministic algorithms with the
following guarantees, for every polygonal curve $H$ (the \emph{host})
of size $b$ and every $r>0$.
\begin{itemize}
\item $\Build(H,r)$ returns a self-index for $H$ and its reversal in
$\tO(b^{1+\alpha})$ time. Its stored boundary information and witness
data are independent of subsequent query labels.
\item $\SelfQuery(U,V,\lambda)$ takes two subcurves $U,V$ of $H$ or of
its reversal, with arbitrary real endpoints, and a labelling
$\lambda$ of $\In(U\times V)$. In $\tO(b^{1+\delta})$ time, including
the report of a local witness for any one retained record, it computes
the min- or max-propagation of $\lambda$ through a boundary map of
$U\times V$ with raw factor at most $D-4$
(Definition~\ref{def:boundary-map}). This map is independent of
$\lambda$, and every record created while applying it
has a local witness of cost at most $Dr$
(Lemma~\ref{fd:rawreports}). In particular, the output is an
$(r,D)$-certified propagation of $\lambda$, represented by output
label profiles with local witness records.
\end{itemize}
\end{definition}

The same boundary map can be used with different input labels. Its
form also gives a compact representation of outputs on the opposite
side (Lemma~\ref{fd:opposite-records}), which we need when mapping
outputs back from the surrogate.

The self-query bound is charged to the host size. To use smaller hosts
inside the recursion, we explicitly prepare additional indexes at
dyadic scales. These indexes still belong to the same outer round.

\begin{definition}[Window bundle]
\label{def:bundle}
Fix an index type and a finite set $\Lambda\subset[1,\infty)$. The
\emph{window bundle} of a curve $H$ at radius $r$ consists of the
indexes $\Build(W,\rho r)$ for every $\rho\in\Lambda$ and every window
$W$ of $H$ that is either an aligned dyadic window
$H[j2^\ell,(j+1)2^\ell]$ or the union of two adjacent aligned dyadic
windows of the same length.
The scales run from one edge to the first power of two at least the
length of the host's parameter interval, with windows truncated at its
endpoints. A one-point host has itself as its only window.
\end{definition}

\begin{lemma}[Window bundles]
\label{ix:bundles}
For an index of type $(\alpha,\delta;D)$, the window bundle of a curve
of size $b$ can be built in $\tO(|\Lambda|\,b^{1+\alpha})$ time. Every
subcurve of size $w$ is contained in a window of the bundle of size
$O(w)$. Consequently, a self-query on subcurves $U,V$ that lie in a
common window of size $O(w)$, with either orientation allowed on each
axis, takes $\tO(w^{1+\delta})$ time.
\end{lemma}
\begin{proof}
At scale $2^\ell$, there are $O(b/2^\ell+1)$ windows and adjacent
unions, each of size $O(2^\ell)$, so their indexes cost
$\tO(b\,2^{\ell\alpha})=\tO(b^{1+\alpha})$ per radius. Summing over the
$O(\log b)$ scales and the radii in $\Lambda$ gives the first claim. A
subcurve of size $w$ meets at most two consecutive aligned windows of
length $2^\ell$, where $w\le2^\ell<2w$, and their union is a prepared
window of size $O(w)$.
\end{proof}

The base index has type $(1,0;9)$, with construction and analysis
given in Section~\ref{ix:section}.

We can now state the query guarantee. The radius multipliers below
specify the indexes needed by its self-queries.

\begin{theorem}[Inner recursion]
\label{qr:query}
Fix an index of type $(\alpha,\delta;D)$ with $\delta<1$, and define
\begin{equation}
 \beta_0=1,\qquad
 \beta_{t+1}=\frac{\beta_t(1+\delta)}{1+\beta_t},\qquad
 A_0=1,\qquad
 A_{t+1}=D(A_t+1)+A_t.
 \label{eq:framework-inner}
\end{equation}
Then $A_t=2(D+1)^t-1$. Let $t\ge0$ and suppose that window bundles
with radius multipliers $\Lambda_t=\{1\}\cup\{A_j+1:0\le j<t\}$ are
available separately on $P$ and $Q$. For subcurves $P',Q'$ of sizes
$p,q$ and a labelling $\lambda$ of $\In(P'\times Q')$, a deterministic
algorithm $\Query_t(P',Q',\lambda)$ returns output label profiles
forming an $(r,A_t)$-certified min- or max-propagation of $\lambda$ in
$\tO((p+q)\min\{p,q\}^{\beta_t})$ time, including the report of one
local witness from an input point of $P'\times Q'$. It uses no joint
index on $P$ and $Q$, even when they lie on a common host. Its search
choices and the boundary relations used for propagation are
independent of $\lambda$.
\end{theorem}

Search tests use radius $r$, while depth $j+1$ uses self-queries at
radius $(A_j+1)r$. Thus only the multipliers in $\Lambda_t$ are needed.
We prove the theorem by induction. The base routine $\Query_0$ is
exact propagation (Lemma~\ref{fd:cell}), taking $\tO(pq)$ time without
an index. Suppose that $\Query_t$ is available, and put
$\beta=\beta_t$ and $A=A_t$. For the recursive step, let
$R=P'\times Q'=[x_0,x_1]\times[y_0,y_1]$, where
$p=|P'|\le b$ and $q=|Q'|\le b$, and set $\rho=(A+1)r$.
We first obtain a bound in terms of $b$, then handle unbalanced
rectangles by subdivision. Rectangles of zero width or height use
exact propagation in near-linear time at every depth.

Algorithm~\ref{alg:query-overview} combines two components. The
\emph{cursor search} constructs a surrogate using fresh, equal labels,
independently of the caller's labels. The \emph{surrogate query} then
propagates the caller's labels through its donors and joins. We give
the two routines below for bottom sources and min-propagation.
Left sources are handled by exchanging the axes, and max-propagation
is symmetric. We first restrict the caller's labels to the $r$-free
input boundary and close them forward there. This preserves every
$r$-reachable pair. Write $\lambda_{\rm bot}$ for the resulting
bottom labels.

\subsection{The cursor search}

The search selects donors without using the caller's labels. Every
test starts from all free bottom points, so one search serves all
sources. Failed tests determine how far the cursor can advance without
losing an $r$-reachable output.

For $1\le k\le b$, split $Q'$ at
$y_0=c_0<_{Q'}c_1<_{Q'}\cdots<_{Q'}c_{k'}=y_1$ into
$k'\le\min\{k,q\}$ chunks, each of size at most $H=O(\lceil b/k\rceil)$.
The internal cuts are at original vertices.
A \emph{test} for chunk $j$ with left end $\kappa$ and width $W$ runs
$\Query_t$ at radius $r$ on the rectangle
$P'[\kappa,\min\{\kappa+W,x_1\}]\times Q'[c_{j-1},c_j]$, with every
free point of its bottom side as a source of label $0$ and unavailable
labels elsewhere. The test
\emph{succeeds} if some point of its top side receives a finite label.
The search starts at $\kappa_0=x_0$ and doubles the tested width until
it succeeds or reaches the end of $P'$. Algorithm~\ref{alg:cursor-search}
keeps the right endpoint $L$ of the last failed test. After a success,
it advances to the later of $L$ and the donor endpoint. We write
$\kappa_j$ for this updated cursor and $U_j=P'[u_j,v_j]$ for the donor.
If the last test fails, a right output may still supply a final
partial donor. Otherwise the represented prefix ends at $c_{j-1}$.
Let $\ell$ be the number of donors. The extreme accepted points used
below are attained because the finite parts of the output profiles
are closed.

\begin{algorithm}[!htb]
\caption{$\mathsf{CursorSearch}_t(P',Q',k,r)$}
\label{alg:cursor-search}
\begin{algorithmic}[1]
\Require Separate window bundles for $P,Q$ and the routine $\Query_t$
\Ensure A surrogate $S$, its matching $\mu$ with $Q'[y_0,y^*]$, and $y^*$, or an empty result
\State Partition $Q'$ into chunks $Q'[c_{j-1},c_j]$ of size $O(\lceil b/k\rceil)$
\State Initialize $\kappa\gets x_0$, $y^*\gets y_0$, and an empty donor list
\For{each chunk $j$, in order}
 \State $L\gets\kappa$
 \For{$W=1,2,4,\ldots$, stopping when the tested prefix reaches $x_1$}
  \State $e\gets\min\{\kappa+W,x_1\}$
  \State Run $\Query_t$ on $P'[\kappa,e]\times Q'[c_{j-1},c_j]$ at radius $r$,
   with label $0$ on all free bottom points and unavailable labels elsewhere
  \If{a top output is finite}
   \State Report a witness $\omega_j$ to the leftmost such point $(v_j,c_j)$,
    starting at $(u_j,c_{j-1})$
   \State Append donor $U_j=P'[u_j,v_j]$ with witness $\omega_j$ and set $y^*\gets c_j$
   \State Set $\kappa\gets\max\{v_j,L\}$ and \textbf{break} the width loop
  \ElsIf{$e=x_1$}
   \If{a right output is finite}
    \State Let $(x_1,y^*)$ be the highest such point and report its witness $\omega_j$
     from $(u_j,c_{j-1})$
    \State Append the partial donor $U_j=P'[u_j,x_1]$ with witness $\omega_j$
   \EndIf
   \State \textbf{Break} both loops
  \Else
   \State $L\gets e$
  \EndIf
 \EndFor
\EndFor
\If{no donor was found}
 \State \Return $(\varnothing,\varnothing,y_0)$
\EndIf
\State Join consecutive donors to form $S$ and combine their witnesses into $\mu$
 \Comment{Lemma~\ref{qr:surrogate}}
\State \Return $(S,\mu,y^*)$, retaining the donors as the pieces of $S$
\end{algorithmic}
\end{algorithm}

An approximate test may accept a donor ending inside an earlier failed
prefix. Updating the cursor to $\max\{v_j,L\}$ still charges large
tests to forward progress. The next lemma proves both this time bound
and the coverage needed for completeness.

\begin{lemma}[Cursor search]
\label{qr:cursor}
Algorithm~\ref{alg:cursor-search} has the following guarantees.
\begin{enumerate}
\item $u_1\le_{P'}v_1\le_{P'}u_2\le_{P'}v_2\le_{P'}\cdots$.
Each $\omega_j$ is an $Ar$-matching of $U_j$ and its represented chunk.
Together, the straightened witnesses have $O(p+q+k)$ segments.
\item For every bottom point $s$ and output point $z$,
$s\Rr r z$ implies $\ell\ge1$ and $z_y\le_{Q'}y^*$.
In particular, a reachable top point implies $y^*=y_1$.
\item The total search time, including witness reporting, is
$\tO(bH^\beta)$.
\end{enumerate}
\end{lemma}
\begin{proof}
For (1), $u_j\ge_{P'}\kappa_{j-1}\ge_{P'}v_{j-1}$ for $j\ge2$, and
soundness of $\Query_t$ gives $\omega_j$. The donors have disjoint interiors, so
$\sum_j(|U_j|+|Q'[c_{j-1},\min\{c_j,y^*\}]|)=O(p+q+k)$.
Straightening within each visited cell gives the same bound on the
number of witness segments.

For (2), fix an $r$-path $\pi:s\to z$. For every cut $c_j\le_{Q'}z_y$,
let $\xi_j=\min\{x:(x,c_j)\in\pi\}$, so $\xi_0=s_x$.
Starting from $\kappa_0=x_0\le_{P'}\xi_0$, we inductively maintain
$\kappa_j\le_{P'}\xi_j$.
If $\kappa_{j-1}\le_{P'}\xi_{j-1}$, every test ending at
$e\ge_{P'}\xi_j$ contains the portion of $\pi$ between these cuts.
Every such test accepts $\xi_j$ by completeness, so failed tests end
before $\xi_j$. A successful test either ends before $\xi_j$ or
accepts it, giving
\[
 L\le_{P'}\xi_j,\qquad v_j\le_{P'}\xi_j,\qquad
 \kappa_j=\max\{v_j,L\}\le_{P'}\xi_j,
\]
where $L=\kappa_{j-1}$ if no test failed.
Thus all chunks crossed by $\pi$ succeed. For a right output $z$,
choose the first $j$ with $z_y\le_{Q'}c_j$. Either chunk $j$ succeeds,
or its final test contains the remaining portion of $\pi$ and accepts
$z$ on the right. In both cases $y^*\ge_{Q'}z_y$ and a donor exists.
This also covers $z_y=y_0$, when the remaining portion lies on the
bottom boundary.

For (3), let $\mathcal W_j$ be the tested widths after clipping at
$x_1$, let $W_j=\max\mathcal W_j$, and for successful chunks put
$d_j=\kappa_j-\kappa_{j-1}$. A success after a failure has
$d_j\ge W_j/2$. A success without a failure has $W_j\le1$, and the
final unsuccessful search has $W_j\le p$. Hence
\[
 \sum_j W_j\le 2\sum_{j\text{ successful}}d_j+k+p=O(p+k).
\]
Doubling gives $\sum_{W\in\mathcal W_j}W=O(W_j)$ and
$|\mathcal W_j|=O(\log(b+2))$. By the inductive query bound and
$k'H=O(b)$, the total time is
\[
 \sum_j\sum_{W\in\mathcal W_j}\tO((W+H)H^\beta)
 =\tO\!\left(H^\beta\Bigl(\sum_jW_j+k'H\Bigr)\right)
 =\tO(bH^\beta).
\]
Each successful or partial chunk reports one witness within its test's
time bound.
\end{proof}

\subsection{Propagation through the surrogate}

The donors give pieces of $P'$ that match consecutive chunks of the
represented prefix $Q'[y_0,y^*]$. Joining them yields a surrogate for
this prefix within distance $Ar$.

\begin{lemma}[Surrogate]
\label{qr:surrogate}
Let $S$ be the curve formed by the donors $U_1,\ldots,U_\ell$, $\ell\ge1$, in order,
where consecutive donors are joined by the segment from $P(v_j)$ to
$P(u_{j+1})$. Then the witnesses $\omega_j$, together with the joining
segments matched to $Q'(c_j)$ while $Q'$ pauses, form an $Ar$-matching
$\mu$ of $S$ and $Q'[y_0,y^*]$ with $O(p+q+k)$ segments. Both $S$ and
$\mu$ can be constructed in $O(p+q+k)$ time.
\end{lemma}
\begin{proof}
For each join $ac$, the adjacent donor witnesses give
$\max\{\|a-Q'(c_j)\|,\|c-Q'(c_j)\|\}\le Ar$.
By convexity, every point of $ac$ is within $Ar$ of $Q'(c_j)$.
Concatenation gives $\mu$. There are $O(p+q+k)$ witness segments and
$O(k)$ joins, so this construction takes $O(p+q+k)$ time.
\end{proof}

We next propagate the input labels through $P'\times S$. Both axes
of a donor rectangle $P'\times U_j$ lie in a common prepared window
of size $O(b)$, so it is handled by $\SelfQuery$ at radius $\rho$
(Lemma~\ref{ix:bundles}). The first donor starts within $Ar$ of
$Q'(y_0)$. Thus every $r$-free bottom source remains $\rho$-free when
its label is placed on the first donor rectangle.

Joins need separate treatment because they need not follow the host.
Across a join $ac$, we hold $P'$ at a position $x$ and carry its label
forward if both $a$ and $c$ are within $\rho$ of $P(x)$. The test is
an interval restriction on each edge of $P'$. It suffices for the
original paths: at a common chunk endpoint, the original matching and
both donor endpoints are close to the same point of $Q'$. The final
completeness proof makes this synchronization explicit.

\begin{algorithm}[!htb]
\caption{$\mathsf{JoinTransfer}(P',a,c,\phi,\rho)$}
\label{alg:join-transfer}
\begin{algorithmic}[1]
\Require A join $ac$ and the top profile $\phi$ of the preceding donor rectangle
\Ensure Bottom labels for the next donor rectangle, with join witnesses
\State Initialize the returned labelling $\psi$ to be unavailable
\For{each closed edge interval $I$ of $P'$}
 \State Compute $J_I=\{x\in I:\|P(x)-a\|\le\rho,\ \|P(x)-c\|\le\rho\}$
 \State Set $\psi|_{J_I}\gets\phi|_{J_I}$, retaining a predecessor pointer to $\phi$
 \State Attach the witness that holds $P'$ at $P(x)$ while traversing $ac$
\EndFor
\State Apply the same endpoint tests to the vertex labels, including the ends of $P'$
\State \Return $\psi$
\end{algorithmic}
\end{algorithm}

\begin{lemma}[Join transfer]
\label{qr:join}
Let $J=\{x:\max\{\|P(x)-a\|,\|P(x)-c\|\}\le\rho\}$.
Algorithm~\ref{alg:join-transfer} returns $\psi=\phi$ on $J$ and
unavailable labels elsewhere, in $O(p\log M)$ time. For every $x\in J$,
traversing $ac$ while holding $P'$ at $P(x)$ has cost at most $\rho$.
Conversely, every traversal of $ac$ at a fixed $x$ and cost
at most $\rho$ has $x\in J$.
\end{lemma}
\begin{proof}
For each edge interval $I$, $J\cap I$ is the intersection of two
free intervals and is computed in $O(1)$ time. Lemma~\ref{fd:tree}
restricts its profile in $O(\log M)$ time. There are $O(p)$ edges
and vertices. For $x\in J$ and $0\le\theta\le1$,
\[
 \|P(x)-((1-\theta)a+\theta c)\|
 \le(1-\theta)\|P(x)-a\|+\theta\|P(x)-c\|\le\rho.
\]
The converse follows from the two endpoints.
\end{proof}

The restriction returned by $\mathsf{JoinTransfer}$ need not be
monotone on the next free edge. The next self-query first applies
input closure, implemented by Lemma~\ref{fd:boundary-assembly}, in
$O(b\log M)$ time. Thus each join supplies valid input profiles
without reading their individual steps.

\begin{algorithm}[!htb]
\caption{$\mathsf{SurrogateQuery}(P',S,\lambda_{\rm bot},\rho)$}
\label{alg:surrogate-query}
\begin{algorithmic}[1]
\Require Ordered donors $U_1,\ldots,U_\ell$ with $\ell\ge1$, an index at radius $\rho$, and
 bottom labels $\lambda_{\rm bot}$ supported on $\rho$-free points
\Ensure Right-output records of the donor rectangles and the final top profile
\State Initialize $\lambda_1$ from $\lambda_{\rm bot}$ on the bottom of $P'\times U_1$,
 with unavailable labels elsewhere
\For{$j=1,\ldots,\ell$}
 \State $\eta_j\gets\SelfQuery(P',U_j,\lambda_j)$ at radius $\rho$
 \State Save the right-output records of $\eta_j$ and their predecessor pointers
 \If{$j<\ell$}
  \State Let $a$ be the end of $U_j$ and $c$ the start of $U_{j+1}$
  \State $\psi\gets\mathsf{JoinTransfer}(P',a,c,\eta_j|_{\rm top},\rho)$
  \State Initialize $\lambda_{j+1}$ from $\psi$ on the bottom of $P'\times U_{j+1}$,
   with unavailable labels elsewhere
 \EndIf
\EndFor
\State \Return the saved records and $\eta_\ell|_{\rm top}$
\end{algorithmic}
\end{algorithm}

Algorithm~\ref{alg:surrogate-query} alternates self-queries with join
transfers, passing all source labels together. Its completeness
guarantee concerns paths that cross each join with $P'$ fixed.

\begin{lemma}[Surrogate query]
\label{qr:transfer}
For $1\le\ell\le k$, Algorithm~\ref{alg:surrogate-query} takes
$\tO(kb^{1+\delta})$ time, including the report of one witness.
Let $\eta$ denote its labels on the donor right sides and the final
top side.
\begin{enumerate}
\item If a $\rho$-matching in $P'\times S$ connects a bottom point $s$
to such an output $z$ and traverses each join it meets at a fixed position on
$P'$, then $\eta(z)\le\lambda_{\rm bot}(s)$.
\item Every finite label $L$ returned at $z$, or stored in a record
valid at $z$, has a witness $\pi$ from a bottom point $s$ to $z$ with
$\lambda_{\rm bot}(s)=L$ and
$\pi\subseteq(P'\times S)\cap\mathcal F_{D\rho}$.
\end{enumerate}
\end{lemma}
\begin{proof}
For (1), follow the given path through the donor rectangles. Each
self-query and each join preserves the inequality
$\text{label}\le\lambda_{\rm bot}(s)$ by completeness and
Lemma~\ref{qr:join}. Input closure can only decrease labels.

For (2), traceback gives local paths $\pi_1,\ldots,\pi_a$ through
donors and joins, all retaining the same label. By the index
guarantee and Lemma~\ref{qr:join},
\[
 \operatorname{cost}(\pi_1\cdots\pi_a)
 =\max_h\operatorname{cost}(\pi_h)\le D\rho.
\]
The $\rho$-free boundary portions added by input closure satisfy the
same bound. There are at most $k$ self-queries and $k-1$ joins, so
the running time is $k\,\tO(b^{1+\delta})+O(kb\log M)
=\tO(kb^{1+\delta})$. A witness uses at most one local witness per
donor and one traversal per join, within this bound.
\end{proof}

It remains to return the outputs from $P'\times S$ to $R$.
The top outputs retain their $P'$-coordinates. Right outputs change
axes through $\mu$, so we use the scalar records of
Lemma~\ref{fd:opposite-records}. Keeping these records avoids
reparameterizing a long profile. Each record has its own witness,
whether or not it attains the minimum at a given point.
To recover a witness on the original curves, we compose matchings
through their common curve. The costs add, and merging the
breakpoints along that curve takes linear time.

\begin{lemma}[Output transfer]
\label{qr:records}
For a right-output record $(L,J)$ on a donor, assign label $L$ to
\[
 \mu(J)=\{y:(u,y)\in\mu\text{ for some }u\in J\}
\]
on the right side of $R$. If $y^*=y_1$, also copy the final top profile
with its $P'$-coordinates unchanged. Restrict these contributions to
the $r$-free output boundary and close them forward.
The resulting profiles can be computed in
$O((p+q+k)\log M)$ time. Every finite label has a witness from a
bottom source carrying that label, of cost at most $D\rho+Ar$.
The same statements hold for max-propagation.
\end{lemma}
\begin{proof}
By Lemma~\ref{fd:opposite-records}, each donor edge contributes
$O(1)$ records. Including singleton outputs of point donors, there
are $O(p+k)$ records in total. For $J=[u^-,u^+]$,
monotonicity of $\mu$ gives
\[
 \mu(J)=\left[\min\{y:(u^-,y)\in\mu\},
                   \max\{y:(u^+,y)\in\mu\}\right].
\]
Scan the record endpoints, the segments of $\mu$, and the vertices of
$Q'$ in order. Only $O(1)$ records cover any point of a donor edge,
so splitting their images at the vertices of $Q'$ produces
$O(p+q+k)$ intervals in total. The scan takes the same time.
Lemma~\ref{fd:boundary-assembly} then restricts and closes the scalar
records and the inherited top profiles in $O((p+q+k)\log M)$ time.

For a mapped label at $(x_1,y)$, choose $u\in J$ with $(u,y)\in\mu$.
Lemma~\ref{qr:transfer} supplies a matching of $P'[s_x,x_1]$ and
$S[0,u]$ of cost at most $D\rho$, while the prefix of $\mu$ matches
$S[0,u]$ to $Q'[y_0,y]$ at cost at most $Ar$. Composing these
matchings through $S[0,u]$ gives a witness of cost at most $D\rho+Ar$.
For top outputs, compose with all of $\mu$ instead. Output closure
adds only $r$-free boundary portions and preserves this bound.
\end{proof}

Finally, run the same construction for left sources. On each output
edge, one answer contributes an inherited profile and the other
contributes scalar records, which we retain for this final merge.
There are $O(p+q+k)$ scalar records in
total, so Lemma~\ref{fd:boundary-assembly} combines the two answers
within the output-transfer bound.

\subsection{Completing the recursion}

The remaining argument connects these components to the original
reachability problem. For completeness, compose an original path with
$\mu$ through $Q'$. For soundness, compose a reported path with $\mu$
through $S$.

\begin{proof}[Proof of Theorem~\ref{qr:query}]
The base case follows from Lemma~\ref{fd:cell}, since
$pq\le(p+q)\min\{p,q\}$. Assume the theorem for $\Query_t$, and use
$\beta=\beta_t$, $A=A_t$, and $\rho=(A+1)r$ as above.

\emph{Completeness.}
Let $\pi:s\to z$ be an $r$-path in $R$, with $s$ on the bottom side.
Lemma~\ref{qr:cursor} gives $\ell\ge1$ and $z_y\le_{Q'}y^*$.
Choose a donor position $u$ with $(u,z_y)\in\mu$, taking the end of
$S$ if $z$ is on the top side. Compose $\pi$ with the prefix of $\mu$
ending at $(u,z_y)$ through their common curve $Q'[y_0,z_y]$.
At each pause on $Q'$, advance along $P'$ first and then along $S$.
This yields a path $\sigma:(s_x,0)\to(z_x,u)$ in $P'\times S$.
In particular, $P'$ stays fixed while $S$ traverses a join at a cut
$c_j$, and every point of this traversal satisfies
\[
 \|P'(x)-S(v)\|
 \le\|P'(x)-Q'(c_j)\|+\|Q'(c_j)-S(v)\|
 \le r+Ar=\rho.
\]
The same triangle inequality bounds the rest of $\sigma$ by $\rho$.
Lemma~\ref{qr:transfer} therefore gives
$\eta(z_x,u)\le\lambda_{\rm bot}(s)$.
For a right output, $\mu$ carries this contribution to $z$.
For a top output, $y^*=y_1$ and its horizontal coordinate is unchanged.
Since $z$ is $r$-free, the final restriction preserves the contribution
and output closure can only decrease its label. Exchanging the axes
proves the claim for left sources. Taking the minimum of both answers
proves completeness.

\emph{Soundness and approximation.}
By Lemmas~\ref{qr:transfer} and~\ref{qr:records}, each finite output
retains its source label and has a witness of cost at most
\[
 D\rho+Ar=\bigl(D(A_t+1)+A_t\bigr)r=A_{t+1}r.
\]
The initial input closure and final output closure add only $r$-free
boundary portions. Their predecessor records recover a source carrying
the same original input label. Moreover,
$A_{t+1}+1=(D+1)(A_t+1)$ and $A_0=1$, whence
$A_t=2(D+1)^t-1$.

\emph{Running time.}
Lemmas~\ref{qr:cursor}, \ref{qr:transfer}, and~\ref{qr:records} give,
for both input sides together,
\[
 \tO\bigl(b(b/k)^\beta+kb^{1+\delta}+b\bigr)
 =\tO\bigl(b(b/k)^\beta+kb^{1+\delta}\bigr).
\]
Inductively $\beta>\delta$, since
$\beta_{t+1}-\delta=(\beta_t-\delta)/(1+\beta_t)>0$ and $\beta_0=1$.
Choose $k=\lceil b^\gamma\rceil$, where
$\gamma=(\beta-\delta)/(1+\beta)\in(0,1)$. Then
\[
 1+\beta(1-\gamma)=1+\delta+\gamma
 =1+\frac{\beta(1+\delta)}{1+\beta}=1+\beta_{t+1}.
\]
If fewer than $k$ chunks are possible, use single-edge chunks.
Their number and size remain within the bounds used above.
Reporting a witness visits at most one local witness per donor.
Composing it with $\mu$ takes time linear in their total size, which
is within $\tO(kb^{1+\delta})$.
The search calls use $\Lambda_t$, and the new self-queries use
$A_t+1$, so $\Lambda_{t+1}=\Lambda_t\cup\{A_t+1\}$ suffices.
The search choices, join intervals, and boundary maps are independent
of the caller's labels.

\emph{Unbalanced rectangles.}
For $p\ge q$, partition $P'$ into $O(1+p/q)$ subcurves of size $O(q)$.
Process their rectangles with $Q'$ from left to right, supplying the
original bottom labels and the original left labels to the first
rectangle. Pass each right output to the next left input.
Lemma~\ref{ix:dag} preserves the factor $A_{t+1}$, and the
window bundles provide hosts of size $O(q)$. The total time, including
one witness, is
\[
 \tO\bigl((1+p/q)q^{1+\beta_{t+1}}\bigr)
 =\tO\bigl((p+q)\min\{p,q\}^{\beta_{t+1}}\bigr).
\]
The case $p<q$ is symmetric. This construction of $\Query_{t+1}$
uses only the unbalanced $\Query_t$, so the induction is well founded.
\end{proof}

The recurrence shows precisely how deeper inner recursion approaches
the self-query bound. Solving it gives the following expressions,
which we use to choose a common depth in the outer recursion.

\begin{lemma}[Query exponents]
\label{qr:exponents}
For $\beta_t$ in~\eqref{eq:framework-inner},
$\beta_t=1/(t+1)$ if $\delta=0$, and
$\beta_t=\delta/(1-(1-\delta)(1+\delta)^{-t})$ if $0<\delta<1$.
In both cases, $\beta_t>\delta$ and $\beta_t\downarrow\delta$.
\end{lemma}
\begin{proof}
Set $z_t=1/\beta_t$. Then $z_0=1$ and
$z_{t+1}=(z_t+1)/(1+\delta)$. Hence $z_t=t+1$ for $\delta=0$.
Otherwise,
$z_t=\delta^{-1}+(1-\delta^{-1})(1+\delta)^{-t}\uparrow\delta^{-1}$.
Taking reciprocals proves the claims.
\end{proof}

%% file: sections/06-index-construction.tex

\section{The outer recursion: constructing indexes}
\label{sec:index}

An outer round uses fast general queries on short hosts to build a
new self-index. We prove the preprocessing and self-query bounds of
Section~\ref{intro:index}, starting with the base index.
Both constructions store maps of canonical rectangles and combine
these maps to answer a query. Before applying the caller's labels,
we compile the resulting network into one boundary map of the whole
query rectangle (Lemma~\ref{ix:compiler}). This gives the interface
of Definition~\ref{def:index}, including the compact opposite-side
records needed by the inner recursion.

\subsection{The base index}
\label{ix:section}

Let $H$ be a host of size $b\ge2$. Its dyadic intervals are the subcurves
$H[j2^\ell,(j+1)2^\ell]$, truncated at the host endpoints, for
$0\le\ell\le\lceil\log_2(b-1)\rceil$. We store maps for every ordered
pair of these intervals, with both orientations allowed on each
axis. Each map is compiled from exact propagation, so its raw factor
is $1$ and its factor is $5$.

For a self-query on $U\times V$, separate the at most two partial
original edges at the ends of each axis. Cover the remaining whole
edges by maximal dyadic intervals. The products of these intervals
have stored maps, while the remaining strips are handled by exact
propagation. We compile network propagation through this partition
into a map of $U\times V$, then apply the caller's labels.
One-point hosts and degenerate query rectangles use exact propagation
directly, followed by the same compilation.

\begin{lemma}[Base index]
\label{ix:base-index}
The routines above form an index of type $(1,0;9)$.
\end{lemma}
\begin{proof}
For scales $a=2^\ell$ and $c=2^{\ell'}$, there are
$O((b/a)(b/c))$ products, each compiled in $\tO(ac)$ time by
Lemmas~\ref{fd:cell} and~\ref{ix:compiler}. Hence
\[
 T_{\rm build}
 =\sum_{\ell,\ell'}O\!\left(\frac ba\frac bc\right)\tO(ac)
 =\tO(b^2).
\]
Let $\mathcal U,\mathcal V$ be the dyadic covers of the whole-edge
parts of a query. They have $O(\log(b+2))$ intervals each, and
\[
 \sum_{I\in\mathcal U,\,J\in\mathcal V}(|I|+|J|)
 =|\mathcal V|\sum_{I\in\mathcal U}|I|
  +|\mathcal U|\sum_{J\in\mathcal V}|J|
 =O(b\log(b+2)).
\]
The partial-edge strips meet $O(b)$ cells in total. Applying the
stored maps and propagating through the strips therefore takes
$\tO(b)$ time. A stored-map witness for $I\times J$ is reported in
$\tO(|I|+|J|)$ time, again by Lemmas~\ref{fd:cell}
and~\ref{ix:compiler}. Thus the network is $(r,5)$-certified, with
$\tau_{\rm run},\tau_{\rm wit}=\tO(b)$
(Lemma~\ref{ix:dag}). Compiling this network gives a map with raw
factor $5$ and factor $9$. Applying it and reporting a witness for
any retained record take $\tO(b)$ time. Degenerate cases have the same
bounds by exact propagation.
\end{proof}

\subsection{Preprocessing on shorter hosts}

We now replace the quadratic preprocessing by calls to the old index
on short blocks. General queries construct the connections between
blocks, and the resulting primitive maps supply all larger canonical
maps. The following theorem gives one outer step.

\begin{theorem}[Outer recursion]
\label{ix:upgrade}
Fix an index of type $(\alpha,\delta;D)$, an integer $T\ge0$, and a
real number $\beta$ with $\beta_T\le\beta<\alpha$, where $\beta_T$ and
$A_T$ are given by~\eqref{eq:framework-inner}. Then there is an index
of type
\begin{equation}
 \Bigl(\frac{\alpha}{1+\alpha-\beta},
       \frac{\beta}{1+\alpha-\beta};\ A_T+12\Bigr).
 \label{eq:framework-outer}
\end{equation}
Its construction on a host of size $b$ invokes the given index only on
hosts of size $O(b^{1/(1+\alpha-\beta)})$.
\end{theorem}

Fix the given index and $T$, and write $A=A_T$. We first allow an
arbitrary integer block size $1\le h\le b$, then choose $h$ to balance
the preprocessing costs. Split the host into a family $\mathcal B$ of
$N=O(b/h)$ blocks, each containing at most $h$ consecutive original
edges. A \emph{canonical interval} is an aligned group of $2^\ell$
blocks, for $0\le\ell\le\lceil\log_2N\rceil$, truncated at the end
of the block sequence.
These intervals form a dyadic tree of height $O(\log(N+2))$.
We also use their reversals, so each pair of intervals has four
possible orientations. A one-point host is handled directly as in
the base construction.

On each block, prepare the window bundle of the given index for the
radius multipliers $\Lambda_T$ of Theorem~\ref{qr:query}. Collect
these bundles in $\mathcal J$. They include indexes on shorter windows
within the block, all at the same old outer level. Thus a call to
$\Query_T$ on subcurves of two blocks uses indexes at the required
local scales, even when the subcurves have arbitrary real endpoints.
The subcurves themselves need not be indexed windows.

The routine $\mathsf{StoreMaps}$ first compiles $\Query_T$ on each
block pair into a \emph{primitive map}. For a canonical rectangle,
it then compiles network propagation through the primitive maps inside
that rectangle. Each canonical map is built directly from primitive
maps, so the approximation loss does not grow with the number of
scales. This implements the last step of
Algorithm~\ref{alg:build-overview}, with the outer-round subscript
omitted.

\begin{algorithm}[!htb]
\caption{$\mathsf{StoreMaps}(\mathcal B,\mathcal J,r)$}
\label{alg:store-maps}
\begin{algorithmic}[1]
\Require Blocks $\mathcal B$, their old-index bundles $\mathcal J$, and the routine $\Query_T$
\Ensure Primitive and canonical boundary maps, with retained witness states
\For{each ordered pair of oriented blocks $(U,V)$}
 \State Compile $\Query_T(U,V,\cdot)$ into $M^{\rm prim}_{U,V}$
 \Statex \Comment{$\tO(h^{1+\beta})$ time, factor $A+4$}
\EndFor
\For{each ordered pair of oriented canonical intervals $(I,J)$}
 \State Let $u,v$ be the numbers of blocks in $I,J$
 \State Retrieve the $uv$ primitive maps inside $I\times J$ and their reversals
 \State Compile network propagation through these maps into $M^{\rm can}_{I,J}$
 \Statex \Comment{$\tO(uvh)$ time, factor $A+8$}
\EndFor
\State \Return the maps and their compilation states
\end{algorithmic}
\end{algorithm}

The distinction between constructing a map and reporting its witnesses
is important here. Once primitive maps are available, a canonical
map is constructed by propagating labels through them. Their matching
witnesses are expanded only when a query asks for one.

\begin{lemma}[Stored maps]
\label{ix:stored-maps}
Algorithm~\ref{alg:store-maps} takes $\tO(N^2h^{1+\beta})$ time.
Its maps have the following guarantees.
\begin{enumerate}
\item A primitive map has raw factor $A$ and factor $A+4$.
It is applied in $\tO(h)$ time, and a witness for any retained record
is reported in $\tO(h^{1+\beta})$ time.
\item A canonical map on $u\times v$ blocks has raw factor $A+4$
and factor $A+8$. It is applied in $\tO((u+v)h)$ time, and a witness
for any retained record is reported in $\tO((u+v)h^{1+\beta})$ time.
\end{enumerate}
\end{lemma}
\begin{proof}
For a block pair, Theorem~\ref{qr:query} gives factor $A$ and time
$\tO(h^{1+\beta_T})\le\tO(h^{1+\beta})$, including one witness.
Lemma~\ref{ix:compiler} gives (1) and total primitive-map construction
time $\tO(N^2h^{1+\beta})$.

For a canonical rectangle of $u\times v$ blocks, the primitive network
is $(r,A+4)$-certified. By Lemma~\ref{ix:dag},
\[
 \tau_{\rm run}=uv\,\tO(h),\qquad
 \tau_{\rm wit}=\tO((u+v)h^{1+\beta}).
\]
Compiling the network gives (2). At a fixed pair of dyadic levels,
the canonical products partition the block grid, so
$\sum_{I,J}uv=N^2$. Summing the construction costs over all level
pairs yields
\[
 T_{\rm canonical}
 =\sum_{\ell,\ell'}\tO\!\left(h\sum_{I,J}uv\right)
 =\tO(N^2h)\le\tO(N^2h^{1+\beta}).
\]
All four orientation choices change only constant factors.
\end{proof}

\subsection{Self-queries and the trade-off}

A self-query must also handle rectangles whose sides are not canonical
intervals. We use stored maps for the whole-block middle and compute
new maps only near the endpoints. The following decomposition makes
both the boundary-processing cost and the witness-reporting cost
explicit.

Split each query subcurve into its whole-block middle and at most two
partial end blocks. If both endpoints lie inside one block, the
subcurve is a single partial block and its middle is empty. Cover each
middle by its maximal canonical intervals. Their products form a
collection $\mathcal C$. The remaining strips and corners form at
most eight zones. In each zone, cut the other axis at block boundaries
to obtain a collection $\mathcal F$ of \emph{fringe rectangles}.
Every fringe lies in one pair of blocks.

\begin{lemma}[Query decomposition]
\label{ix:decomposition}
Let $R=U\times V$ have positive area, and set $L=\log(N+2)$.
The partition above can be found in $\tO(N)$ time and satisfies
\[
 |\mathcal C|=O(L^2),\qquad |\mathcal F|=O(N),\qquad
 \sum_{I\times J\in\mathcal C}(u_I+u_J)=O(NL),
\]
where $u_I$ denotes the number of blocks in $I$.
Every $F\in\mathcal F$ is contained in a block pair, and
$\sum_{G\in\mathcal C\cup\mathcal F}|G|=O(bL)$.
\end{lemma}
\begin{proof}
Let $\mathcal U,\mathcal V$ be the maximal canonical covers of the
middle intervals. Then $|\mathcal U|,|\mathcal V|=O(L)$ and
$\sum_{I\in\mathcal U}u_I,\sum_{J\in\mathcal V}u_J\le N$. Hence
\[
 \sum_{I\in\mathcal U,\,J\in\mathcal V}(u_I+u_J)
 =|\mathcal V|\sum_{I\in\mathcal U}u_I
  +|\mathcal U|\sum_{J\in\mathcal V}u_J=O(NL).
\]
Each remaining zone has one axis inside a block and is cut into
$O(N)$ fringes of size $O(h)$. Thus $|\mathcal F|=O(N)$ and
\[
 \sum_{G\in\mathcal C\cup\mathcal F}|G|
 =O(hNL)+O(hN)=O(bL).
\]
The covers take $O(L)$ time, and listing the rectangles takes
$O(L^2+N)=\tO(N)$ time.
\end{proof}

Retrieve the stored maps for $\mathcal C$. For each
$F=U_F\times V_F\in\mathcal F$, compile $\Query_T(U_F,V_F,\cdot)$
using the bundles of its two containing blocks. These calls require
no joint index on the blocks, and the window bundles support the
possibly shorter subcurves $U_F,V_F$. Retrieve or construct the maps
with both axes reversed as well, for the reverse-activity run.

The maps form a network on $R$. Order the zones from bottom-left to
top-right, and use the corresponding grid order inside each zone.
Compile network propagation into a single map of $R$, using fresh
labels, then apply the caller's labels. This is
Algorithm~\ref{alg:selfquery-overview}. The last compilation gives
both the required witness records and the opposite-side representation
of Lemma~\ref{fd:opposite-records}.

\begin{lemma}[Self-query]
\label{ix:self-query}
The procedure above constructs a boundary map of $U\times V$ with
raw factor at most $A+8$, independently of the caller's labels.
Applying this map gives $(r,A+12)$-certified propagation in
$\tO(bh^\beta)$ total time, including the report of a local witness
for any one retained record.
\end{lemma}
\begin{proof}
For a degenerate rectangle, compile exact propagation in $\tO(b)$
time, obtaining factor $5\le A+12$. Otherwise, by
Theorem~\ref{qr:query} and Lemmas~\ref{ix:compiler}
and~\ref{ix:decomposition}, the fringe maps have factor $A+4$ and cost
\[
 T_{\rm fringe}=|\mathcal F|\,\tO(h^{1+\beta})
 =\tO(Nh^{1+\beta})=\tO(bh^\beta).
\]
The canonical maps have factor $A+8$. Once these local maps are
available, Lemmas~\ref{fd:profiles} and~\ref{ix:dag} give a network
with factor $A+8$ and running time
$\tau_{\rm run}=\tO(\sum_G|G|)=\tO(b)$.
Boundary transfers inherit a single profile on each open shared
piece and merge only scalar values at junctions.

A network witness visits each local rectangle at most once.
Lemma~\ref{ix:stored-maps} and the decomposition bounds give
\[
 \tau_{\rm wit}
 \le\tO\!\left(h^{1+\beta}
       \Bigl(\sum_{I\times J\in\mathcal C}(u_I+u_J)
                    +|\mathcal F|\Bigr)\right)
 =\tO(Nh^{1+\beta})=\tO(bh^\beta).
\]
Lemma~\ref{ix:compiler}, with $F=A+8$, compiles this network in
$\tO(b)$ time. Its map has raw factor $A+8$ and factor $A+12$.
Application costs $\tO(b)$, and any retained record has a witness
reported in $O(\tau_{\rm wit}+b)=\tO(bh^\beta)$ time.
\end{proof}

Thus the approximation loss comes from three compilations: primitive
maps, canonical maps, and the whole query. Passing through adjacent
rectangles only concatenates witnesses. Neither the number of blocks
nor the number of dyadic levels adds to the factor.

We can now collect the preprocessing and query costs. Preparing the
old indexes becomes more expensive as $h$ grows, while constructing
maps between blocks becomes cheaper.

\begin{lemma}[Trade-off]
\label{ix:tradeoff}
For $1\le h\le b$, the construction has factor $A+12$, preprocessing
time $\tO(bh^\alpha+b^2h^{\beta-1})$, and self-query time
$\tO(bh^\beta)$, including one local witness.
\end{lemma}
\begin{proof}
Since $|\Lambda_T|\le T+1$ is constant, Lemmas~\ref{ix:bundles}
and~\ref{ix:stored-maps} give
\[
 T_{\rm build}
 =\tO\bigl(Nh^{1+\alpha}+N^2h^{1+\beta}\bigr)
 =\tO\bigl(bh^\alpha+b^2h^{\beta-1}\bigr).
\]
The partition and dyadic trees cost $\tO(b)$, within this bound.
The self-query guarantees are Lemma~\ref{ix:self-query}.
\end{proof}

\begin{proof}[Proof of Theorem~\ref{ix:upgrade}]
Set $\gamma=1/(1+\alpha-\beta)$ and choose an integer
$h\asymp b^\gamma$. Since $0\le\beta<\alpha\le1$, $0<\gamma<1$.
Substituting into Lemma~\ref{ix:tradeoff} gives
\[
 T_{\rm build}=\tO\!\left(b^{1+\alpha\gamma}
                    +b^{2-(1-\beta)\gamma}\right)
             =\tO\!\left(b^{1+\frac{\alpha}{1+\alpha-\beta}}\right),
 \qquad
 T_{\rm self}=\tO\!\left(b^{1+\frac{\beta}{1+\alpha-\beta}}\right).
\]
The factor is $A_T+12$, and the old index is invoked only on blocks
and their windows, all of size $O(h)$.
\end{proof}

Section~\ref{sec:complete} chooses a common inner depth $T$ and
iterates this improvement. The base index and
Theorem~\ref{ix:upgrade} then supply all rounds of the construction.

%% file: sections/07-parameters.tex

\section{Proof of the main theorem}
\label{sec:complete}

The outer recursion improves an index whenever the inner query meets
its target exponent. We now choose a common inner depth for all outer
rounds and obtain the final near-linear index. One self-query then
gives the decision procedure, and a search over the radius completes
the proof of Theorem~\ref{thm:main}.

\subsection{Choosing the parameters}

We use $K$ outer rounds with target excess exponent $\theta=1/K$.
To apply Theorem~\ref{ix:upgrade} in every round, the common inner
depth $T$ must satisfy $\beta_{i-1,T}\le\theta<\alpha_{i-1}$.
The following choice ensures this condition and leaves enough slack
in the final exponents to absorb the polylogarithmic factors.

\begin{lemma}[Parameters]
\label{ix:parameters}
For $\eps>0$, let $\bar\eps=\min\{\eps,1\}$ and set
\[
 K=\left\lceil\frac6{\bar\eps}\right\rceil,\qquad
 \theta=\frac1K,\qquad
 T=\left\lceil\frac4\theta\log\frac{16}\theta\right\rceil.
\]
Define $(\alpha_0,\delta_0;D_0)=(1,0;9)$ and, for $1\le i\le K$,
\begin{equation}
 \alpha_i=\frac{\alpha_{i-1}}{1+\alpha_{i-1}-\theta},\qquad
 \delta_i=\frac{\theta}{1+\alpha_{i-1}-\theta},\qquad
 D_i=2(D_{i-1}+1)^T+11.
 \label{eq:complete-schedule}
\end{equation}
Let $\beta_{j,t}$ be the depth-$t$ query exponent for self-query excess
$\delta_j$, as in~\eqref{eq:framework-inner}. Then
\[
 \beta_{i-1,T}\le\theta<\alpha_{i-1}\quad(1\le i\le K),\qquad
 \alpha_K<\frac{\bar\eps}{2},\qquad
 \delta_K<\frac{\bar\eps}{4}.
\]
For $0<\eps\le1$, also
$2D_K\le\exp(\exp(O(\eps^{-1}\log(2/\eps))))$.
\end{lemma}
\begin{proof}
We have $\theta\le\bar\eps/6\le1/6$ and $K\theta=1$.
The recurrence $\alpha_i^{-1}=1+(1-\theta)\alpha_{i-1}^{-1}$ gives
\begin{equation}
 \alpha_i=\frac{\theta}{1-q_i},\qquad q_i=(1-\theta)^{i+1}.
 \label{eq:alpha-closed}
\end{equation}
Thus $1=\alpha_0>\alpha_1>\cdots>\theta$ and
$\delta_i=\theta\alpha_i/\alpha_{i-1}<\theta$ for $i\ge1$.
Since $q_K<e^{-1}$,
\[
 \alpha_K<\frac{\theta}{1-e^{-1}}<2\theta
 \le\frac{\bar\eps}{3}<\frac{\bar\eps}{2},\qquad
 \delta_K<\theta\le\frac{\bar\eps}{6}<\frac{\bar\eps}{4}.
\]

For the common inner depth, $\beta_{0,T}=1/(T+1)\le\theta$ by
Lemma~\ref{qr:exponents}. For $1\le j<K$,
\[
 q_{j-1}=(1-\theta)^j
 \ge e^{-j\theta/(1-\theta)}\ge e^{-1}>\frac18.
\]
Hence $\alpha_{j-1}-\theta=\theta q_{j-1}/(1-q_{j-1})\ge\theta/8$.
Using $\alpha_{j-1}\le1$, we obtain
\[
 \delta_j\ge\frac\theta2,\qquad
 \theta-\delta_j
 =\frac{\theta(\alpha_{j-1}-\theta)}{1+\alpha_{j-1}-\theta}
 \ge\frac{\theta^2}{16}.
\]
Since $\log(1+\delta_j)\ge\delta_j/2\ge\theta/4$, the choice of $T$
gives
\[
 (1+\delta_j)^{-T}
 \le e^{-T\theta/4}\le\frac\theta{16}
 \le\frac{\theta-\delta_j}{\theta(1-\delta_j)}.
\]
By Lemma~\ref{qr:exponents}, this is equivalent to
$\beta_{j,T}=\delta_j/(1-(1-\delta_j)(1+\delta_j)^{-T})\le\theta$.

For the approximation factor, set $E_i=D_i+1$. Then $E_0=10$ and
$E_i=2E_{i-1}^T+12\le4E_{i-1}^T\le E_{i-1}^{T+1}$, so
$\log E_K\le(T+1)^K\log10$. For $0<\eps\le1$,
$K=O(1/\eps)$ and $T=O(\eps^{-1}\log(2/\eps))$, giving
\[
 \log\log(2D_K)\le K\log(T+1)+O(1)
 =O(\eps^{-1}\log(2/\eps)).\qedhere
\]
\end{proof}

All exponents used to choose block sizes and chunk counts are rational
constants for fixed $\eps$. For positive integers $u,v$, a value within
a constant factor of $b^{u/v}$ can be found by doubling and binary
search, comparing $z^v$ with $b^u$. This takes $O(\log b)$ arithmetic
operations, within the stated bounds.

\subsection{The approximation algorithm}

The parameter lemma allows us to build the final index. We next
explain how an index yields a decision procedure, independently of
the particular parameter choice. As in Algorithm~\ref{alg:decide},
join $P$ and $Q$ by one segment to form a common host $H$, preprocess
$H$, and run a self-query on $R=P\times Q$. Only the initial corner
receives a finite input label.

\begin{lemma}[Fixed-threshold decision]
\label{ix:threshold}
For $r>0$, Algorithm~\ref{alg:decide}, using an index of type
$(\alpha,\delta;C)$, is a $C$-approximate decision procedure taking
$\tO(n^{1+\alpha})$ time. Every YES answer reports a matching of $P,Q$
of cost at most $Cr$ with $O(m+n)$ segments.
\end{lemma}
\begin{proof}
Let $s,t$ be the initial and final corners of $R$, and set
$\lambda(s)=0$ and $\lambda=+\infty$ elsewhere. The self-query output
$\lambda'$ satisfies
\[
 \dF(P,Q)\le r\ \Longrightarrow\ s\Rr r t
 \ \Longrightarrow\ \lambda'(t)=0,
\]
whereas $\lambda'(t)=0$ gives a witness
$\pi:s\to t$ with $\pi\subseteq R\cap\mathcal F_{Cr}$.
Thus $\pi$ matches $P$ to $Q$ without using the joining segment.
It visits $O(m+n)$ cells, and straightening within each cell preserves
its cost and leaves $O(m+n)$ segments. Since $|H|=O(n)$ and
$\delta\le\alpha$, preprocessing, the self-query, and witness reporting
take $\tO(n^{1+\alpha}+n^{1+\delta})=\tO(n^{1+\alpha})$ time.
\end{proof}

Applying this lemma to the final index gives the fixed-radius
algorithm. The threshold search of Lemma~\ref{ix:search} then
computes the distance approximation with only
$O(\log\log(n+2))$ decision calls and $O(n)$ additional time.

\begin{proof}[Proof of Theorem~\ref{thm:main}]
Fix $d$ and $\eps$, and choose $K,\theta,T$ as in
Lemma~\ref{ix:parameters}. Starting from the index of
Lemma~\ref{ix:base-index}, apply Theorem~\ref{ix:upgrade} for
$i=1,\ldots,K$ with target $\beta=\theta$.
The inequalities $\beta_{i-1,T}\le\theta<\alpha_{i-1}$ give, by
induction, an index of type $(\alpha_i,\delta_i;D_i)$ at each round,
with the parameters in~\eqref{eq:complete-schedule}.
The required window bundles are included in these bounds, and the set
of nested radius multipliers depends only on $\eps$.

The final index has $\alpha_K<\bar\eps/2$ and
$\delta_K<\bar\eps/4$. Lemma~\ref{ix:threshold} therefore gives a
$D_K$-approximate decision procedure in time
$\tau=\tO(n^{1+\bar\eps/2})$. By Lemma~\ref{ix:search}, it yields
\[
 \dF(P,Q)\le\widehat D\le2D_K\,\dF(P,Q),\qquad
 O\bigl(n+\tau\log\log(n+2)\bigr)=O(n^{1+\eps}).
\]
The last equality holds because the hidden polylogarithmic degree is
constant for fixed $\eps$ and $\eps-\bar\eps/2>0$.
Set $f(1/\eps)=2D_K$. Lemma~\ref{ix:parameters} gives the stated
bound on $f$, and all steps are deterministic. The same guarantees
for the discrete Fr\'echet distance follow from
Theorem~\ref{thm:discrete} in Appendix~\ref{sec:discrete}.
\end{proof}

The search retains a matching along with its distance estimate, so the
algorithm also provides a certificate for the approximation.

\begin{corollary}[Reporting a matching]
\label{cor:matching}
Within the running time of Theorem~\ref{thm:main}, the algorithm
reports a matching of cost at most $\widehat D$. It has $O(m+n)$
segments for $\dF$, and $O(m+n)$ pairs of original vertices for $\ddF$.
\end{corollary}
\begin{proof}
This is the matching returned by Lemma~\ref{ix:search}, using the
witness guarantee of Lemma~\ref{ix:threshold} and its discrete version.
\end{proof}

%% file: sections/08-conclusion.tex

\section{Conclusion}
\label{sec:discussion}

We presented deterministic constant-factor approximation algorithms
for the Fr\'echet distance and the discrete Fr\'echet distance in
$O(n^{1+\eps})$ time for every fixed $\eps>0$ and every fixed dimension.
Our approach combines preprocessing for queries between subcurves of
a single curve with queries between different curves. Self-queries
make general queries faster, and faster general queries reduce the
cost of preprocessing. Applying these two steps recursively gives
near-linear preprocessing and query time.

The main open question is whether the approximation factor can be
improved. The two recursions reduce the running-time exponent but
accumulate approximation losses, giving a doubly exponential bound in
$1/\eps$. Can this dependence be reduced to a polynomial in $1/\eps$,
or even to a constant independent of $\eps$? It also remains open
whether a constant-factor approximation is possible in
$n\operatorname{polylog} n$ time.

%% file: sections/ai-methodology.tex

\section*{AI methodology}
\phantomsection
\label{sec:ai-methodology}
\addcontentsline{toc}{section}{AI methodology}

We began this project in July 2025 and developed an initial framework
for efficiently propagating reachability information, inspired by
Andoni and Shekel Nosatzki~\cite{AN}. Balancing preprocessing and query
costs to obtain near-linear time for general curves remained unresolved.

We began using AI tools in June 2026, but these early attempts gave
near-linear-time constant-factor approximations only in special cases.
In September 2026, GPT~6 Astra helped us resolve the remaining technical
gaps and complete the analysis. We also used Claude
Opus~5.5 and GPT~6 Astra to revise the exposition and proofs. The authors
take full responsibility for the paper.

%% file: sections/A-threshold-search.tex

\section{From decision to approximation}
\label{sec:search}

This appendix proves Lemma~\ref{ix:search}. A search over thresholds
needs a range of candidate values that contains the distance, and the
size of this range must not depend on the coordinates of the input. We
therefore first compute, in linear time, a coarse estimate whose ratio
to the distance is at most $O(m+n)$. A binary search over the
$O(\log(m+n))$ resulting scales then needs only $O(\log\log(m+n))$
calls of the decision procedure. Algorithm~\ref{alg:approximation}
summarizes the procedure. Throughout, $D_*$ denotes $\dF(P,Q)$ or
$\ddF(P,Q)$.

\begin{lemma}[From decision to approximation]
\label{ix:search}
Let $D_*$ be $\dF(P,Q)$ or $\ddF(P,Q)$, and let $C\ge1$. Suppose that
for every $r>0$, a procedure running in time $\tau$ either answers NO,
in which case $D_*>r$, or answers YES and reports a matching of cost
at most $Cr$ with $O(m+n)$ segments or pairs. Then one can compute
$\widehat D$ with $D_*\le\widehat D\le2CD_*$, together with a matching
of cost at most $\widehat D$, in $O(m+n+\tau\log\log(m+n+2))$ time,
even if the answers of the procedure are not monotone in $r$.
\end{lemma}

\begin{algorithm}[!htb]
\caption{The approximation algorithm}
\label{alg:approximation}
\begin{algorithmic}[1]
\Require Nonempty curves $P,Q$ and a $C$-approximate decision procedure
 satisfying Lemma~\ref{ix:search}
\Ensure $\widehat D$ with $D_*\le\widehat D\le2CD_*$, and a matching
 of cost at most $\widehat D$
\State Compute the seed matching and its cost $U$ \Comment{Lemma~\ref{ix:seed}}
\If{$U=0$}
 \Return $(0,\text{seed matching})$
\EndIf
\State Let $M'$ be the least integer with $2^{M'}\ge4(m+n)+1$, and
 let $s_j=U/2^j$
\If{the procedure answers YES at $s_{M'}$}
 \Return $Cs_{M'}$ and its matching
\EndIf
\State $a\gets0$, $b\gets M'$, and $\widehat D\gets U$ with the seed
 matching
\While{$b-a>1$}
 \State $j\gets\lfloor(a+b)/2\rfloor$
 \If{the procedure answers YES at $s_j$}
  \State $a\gets j$, and if $Cs_j<\widehat D$, set $\widehat D\gets
   Cs_j$ with the reported matching
 \Else
  \State $b\gets j$
 \EndIf
\EndWhile
\State \Return $\widehat D$ and its matching
\end{algorithmic}
\end{algorithm}

The coarse estimate comes from a very simple matching, which traverses
both curves at the same arclength speed. If the curves are close, then
matched points of an optimal matching have similar arclength
positions, because every edge of one curve is followed closely by a
portion of the other curve of almost the same length. The total
discrepancy is at most $O(m+n)$ times the distance, and this gives the
following bound.

\begin{lemma}[Seed]
\label{ix:seed}
In $O(m+n)$ time, one can compute a matching of $P$ and $Q$ of cost $U$
with $D_*\le U\le(2(m+n)+1)D_*$ for $\dF$, and a discrete matching
of cost $U$ with $D_*\le U\le(4(m+n)+1)D_*$ for $\ddF$.
In particular, $U=0$ if and only if $D_*=0$.
\end{lemma}
\begin{proof}
We may remove consecutive duplicate vertices. This does not change
$\dF$, and it does not change $\ddF$ either. Indeed, a discrete
matching of the reduced sequences expands to one of the original
sequences by traversing each run of duplicates while the other index
pauses, and a discrete matching of the original sequences projects to
the reduced ones. Neither operation creates a new pair of positions.
If one curve becomes a single point, a linear scan computes $D_*$
exactly, by convexity of the distance along an edge. Otherwise, let
$0=a_0<\cdots<a_{n'}=A$ and $0=b_0<\cdots<b_{m'}=B$ be the cumulative
arclengths at the vertices of $P$ and $Q$, and let $\widetilde P$ and
$\widetilde Q$ be the arclength parameterizations, which are
$1$-Lipschitz.

We first show that matched points have similar arclength positions.
Let $\pi$ be a $z$-matching of $P$ and $Q$, and let $(x,y)\in\pi$ have
arclength positions $a$ and $b$. We claim that $|a-b|\le2(m+n)z$. Cut
$P[0,x]$ at its vertices into at most $n$ straight pieces, and choose,
monotonically along $\pi$, a partner on $Q$ for each cut point. By the
triangle inequality, each piece is at most $2z$ longer than the portion
of $Q$ between the partners of its endpoints, and these portions are
interior-disjoint subcurves of $Q[0,y]$. Summing gives $a\le b+2nz$, and
by symmetry, $b\le a+2mz$. The same argument applies to discrete
matchings, by summing over the steps.

For $\dF$, we match $\widetilde P(\min\{s,A\})$ with
$\widetilde Q(\min\{s,B\})$ for $0\le s\le\max\{A,B\}$, which is a
matching. Let $\pi$ be an optimal matching of cost $D_*$, and fix $s$.
By symmetry, we may assume
$s\le A$. The point $\widetilde P(s)$ has a partner on $\pi$ at some
arclength $b\in[0,B]$ with $|s-b|\le2(m+n)D_*$. Clamping $s$ to $[0,B]$
does not increase its distance to $b$, so
\[
 \|\widetilde P(s)-\widetilde Q(\min\{s,B\})\|
 \le\|\widetilde P(s)-\widetilde Q(b)\|+|b-\min\{s,B\}|
 \le(2(m+n)+1)D_*.
\]
Between consecutive arclength breakpoints, both positions are affine in
$s$, so their distance is convex. Hence $U$ is the maximum over the
$O(m+n)$ breakpoints and is computed by a merge in $O(m+n)$ time.

For the discrete seed, we pair, for every value $s$ in the merged list
of breakpoints, the vertex of $P$ whose arclength is nearest to $s$
with the vertex of $Q$ whose arclength is nearest to $s$, breaking ties
towards the smaller index. Both indices are nondecreasing in $s$. If
the index on $P$ jumped from $i$ to $i+2$ or more between consecutive
values $s<s'$, then $s<a_{i+1}<s'$, which contradicts the choice of
consecutive values. The same holds for $Q$, so after removing repeated
pairs, we obtain a discrete matching. Consider a pair chosen at
$s=a_i$, with $Q$-index $j$, and let $k$ be a partner of $i$ in an
optimal discrete matching. Then $|a_i-b_j|\le|a_i-b_k|\le2(m+n)D_*$,
hence $|b_j-b_k|\le4(m+n)D_*$, and therefore
$\|p_i-q_j\|\le\|p_i-q_k\|+|b_j-b_k|\le(4(m+n)+1)D_*$. Values
$s=b_j$ are handled symmetrically.

In both cases, $U$ is the cost of an actual matching, so $D_*\le U$,
which also shows that $U=0$ if and only if $D_*=0$.
\end{proof}

With the seed, the proof of Lemma~\ref{ix:search} is a binary search
that relies only on the certificate of each individual answer. This is
important because an approximate decision procedure may answer YES at one
threshold and NO at a larger one, as long as both answers are correct.

\begin{proof}[Proof of Lemma~\ref{ix:search}]
We use Algorithm~\ref{alg:approximation}. If $U=0$, the seed is an
exact answer. Otherwise, the choice of $M'$ and Lemma~\ref{ix:seed}
give $s_{M'}\le U/(4(m+n)+1)\le D_*\le U=s_0$. If the procedure answers
YES at $s_{M'}$, its matching certifies $D_*\le Cs_{M'}\le CD_*$.
Otherwise $D_*>s_{M'}$, and the loop maintains the invariant that
$D_*>s_b$ and $D_*\le\widehat D\le Cs_a$, together with a matching of
cost at most $\widehat D$. Initially,
$\widehat D=U=s_0$. A NO answer at $s_j$ certifies $D_*>s_j$, and a YES
answer at $s_j$ comes with a matching of cost at most $Cs_j$, so both
updates preserve the invariant. At termination, $b=a+1$, so
$\widehat D\le Cs_a=2Cs_b<2CD_*$. There are $O(\log M')=O(\log\log
(m+n+2))$ calls, and the seed takes $O(m+n)$ time.
\end{proof}

%% file: sections/B-discrete.tex

\section{The discrete Fr\'echet distance}
\label{sec:discrete}

This appendix adapts the algorithm to the discrete Fr\'echet distance
of the vertex sequences of $P$ and $Q$. The two recursions and their
time bounds carry over unchanged. The discrete setting is in fact
simpler, because the boundary of a rectangle consists of finitely many
vertex pairs. Hence the lift and the persistent profiles of
Sections~\ref{fd:normalization} and~\ref{fd:application} are not
needed, and the approximation factors become smaller.

\begin{theorem}[Discrete version]
\label{thm:discrete}
Theorems~\ref{qr:query} and~\ref{ix:upgrade}, the properties of the
base index (Section~\ref{ix:base}), and Lemma~\ref{ix:threshold} hold
for the discrete Fr\'echet distance, with the same running times and
with approximation factors that are at most those of the corresponding
statements for $\dF$. Consequently,
Theorem~\ref{thm:main} and Corollary~\ref{cor:matching} hold for
$\ddF$.
\end{theorem}

We first describe the setting. A discrete rectangle is a set
$[x_0..x_1]\times[y_0..y_1]$ of index pairs, with side sizes
$p=x_1-x_0+1$, $q=y_1-y_0+1$, and $b=p+q$. Its input boundary
consists of its left column and bottom row, and its output boundary
consists of its top row and right column. A pair $(i,j)$ is
\emph{$r$-free} if $\|p_i-q_j\|\le r$, and $s\Rr r t$ means
that some sequence of free pairs from $s$ to $t$ inside the rectangle
has all its steps in $\{(1,0),(0,1),(1,1)\}$. Labels live on the
$O(b)$ boundary pairs and are stored as explicit arrays. Min- and
max-propagation use the same labels, unavailable values, and local
witness guarantee as in Section~\ref{sec:preliminaries}.

Boundary maps become simpler in this setting. All boundary pairs serve
as ports, so a boundary map is just a finite summary $B$ of the
form~\eqref{fd:finiteformula}, together with the forward closures along
the input and output boundaries, which consist of steps $(1,0)$ and
$(0,1)$. Lemma~\ref{fd:crossing} splices at a grid point, so
Lemma~\ref{fd:completion} applies verbatim. No lift is needed, and
hence a summary with factor $F$ yields a boundary map with factor $F$
rather than $F+4$. Propagation through such a map takes one
range-extremum query per output pair, that is, $O(b\log b)$ time in
total (Lemma~\ref{fd:ports}).

Networks of rectangles also carry over. Partition a discrete rectangle
into discrete rectangles that share boundary rows and columns, and
assign every pair to a rectangle in the half-open sense of the proof
of Lemma~\ref{pre:compose}. If a pair $u$ is assigned to
$[x_0'..x_1']\times[y_0'..y_1']$, then $u_x<_P x_1'$ and $u_y<_Q y_1'$, unless
the rectangle touches the outer boundary. Therefore, the next pair of a
step $(1,0)$, $(0,1)$, or $(1,1)$ lies in the same closed rectangle.
Every step thus lies in a single rectangle, the decomposition in the
proof of Lemma~\ref{ix:dag} applies without inserting intermediate
pairs, and network propagation is certified as before. The exact base
is the dynamic program of Lemma~\ref{fd:cell}, with time $O(pq)$ and
witnesses of length $O(p+q)$.

The interfaces retain their roles: $\Build(H,r)$ returns an index on
the host and its reversal; $\SelfQuery$ propagates labels through a
single map prepared independently of those labels; and $\Query$ uses
the two separately prepared curves. The outputs of both queries are
label arrays with reportable local witnesses. In this appendix, the
index definition uses $D\ge1$, and a self-query map has raw factor at
most $D$, since the continuous lift and its additive $4$ are absent.

The index constructions then carry over. The analysis of the base
index in Section~\ref{ix:section} and the proof of
Lemma~\ref{ix:tradeoff} apply verbatim,
except that every compilation keeps its factor. Hence the base index
has factor $1$, and the upgrade produces factor $A_T$ instead of
$A_T+12$. Both are at most the corresponding factors for $\dF$.

For the inner recursion, we compose discrete matchings through their
common sequence. Let $\gamma$ be a discrete matching of sequences
$X$ and $Y$ of cost $c_1$, and let $\mu$ be a discrete matching of $Z$
and $Y$ of cost $c_2$. We sweep the indices of $Y$, and at each index,
we first advance through the indices of $X$ that are matched to it and
then through those of $Z$. When the index of $Y$ advances, each of
$\gamma$ and $\mu$ advances its other index by zero or one, so after
removing repeated pairs, all steps lie in $\{(1,0),(0,1),(1,1)\}$.
Every visited pair of indices of $X$ and $Z$ is matched to a common
index of $Y$, so the result is a discrete matching of cost at most
$c_1+c_2$.

With this composition, the inner recursion carries over as well. The
cursor search is unchanged, since the proof of Lemma~\ref{qr:cursor}
uses only the completeness and soundness of $\Query_t$ and
monotonicity. The surrogate $S$ is the concatenation of the vertex
sequences of the donors $U_1,\ldots,U_\ell$, where two consecutive
donors that share their endpoint occurrence are identified there. The
last vertex of $U_j$ and the first vertex of $U_{j+1}$ are both within
distance $Ar$ of the vertex $q_{c_j}$, so the witnesses $\omega_j$,
joined by the step $(1,0)$ between these two vertices while $Q'$ stays
at $c_j$, form a discrete $Ar$-matching $\mu$ of $S$ and the
represented prefix of $Q'$.

At a junction, a label at host index $i$ on the top row of
$P'\times U_j$ is passed to the free pairs at host indices $i$ and
$i+1$ on the bottom row of $P'\times U_{j+1}$. These two transfers
correspond to the steps $(0,1)$ and $(1,1)$ in $P'\times S$, and
further steps along $P'$ are handled by the input closure of the next
self-query. The rule is sound, because both transfers are legal steps
between free pairs. It is also complete, because the composition of a
true matching with $\mu$, constructed as above, advances along $P'$
first and then crosses the junction step of $S$ at a fixed host index.
Right outputs are transferred through $\mu$ pair by pair: a pair
$(x_1,u)$ maps to the interval of indices of $Q'$ that $\mu$ matches to
$u$. Since the donors are interior-disjoint, this takes $O(p+q+k)$
time. The proofs of completeness, soundness, and running time in
Section~\ref{sec:query} then carry over verbatim, with the factor
update $A_{t+1}=D(A_t+1)+A_t$ and the same exponents.

Finally, for Lemma~\ref{ix:threshold}, the host is the concatenation
of the vertex sequences of $P$ and $Q$, the only source is the initial
pair, and a discrete matching visits $O(m+n)$ distinct pairs. Together
with the discrete case of Lemma~\ref{ix:seed}, this proves
Theorem~\ref{thm:discrete}.

%% file: sections/C-related-work.tex

\section{Further related work}
\label{sec:related-work}

Godau~\cite{God91} gave an early algorithm for the Fr\'echet
distance in $O((n^2m+nm^2)\log(nm))$ time. Alt and
Godau~\cite{AG95} reduced this to $O(nm\log(nm))$, using an
$O(nm)$-time decision procedure based on free-space reachability.
For the discrete Fr\'echet distance, Eiter and Mannila~\cite{EM94}
gave an $O(nm)$-time dynamic program that computes the distance
directly.
Later exact algorithms improved the logarithmic factors. Agarwal,
Ben Avraham, Kaplan, and Sharir~\cite{AAKS13} obtained
$O(nm\log\log n/\log n)$ time for the discrete Fr\'echet distance
in the plane on a word RAM. Buchin, Buchin, Meulemans, and
Mulzer~\cite{BBMM14} gave planar algorithms for the Fr\'echet
distance on both pointer machines and word RAMs, with expected time
$O(nm(\log\log n)^2)$ on a word RAM. Cheng and Huang~\cite{CH25}
reported a subquadratic bound in arbitrary fixed dimension. For two
$n$-vertex curves, Chan and Hair~\cite{CH26} obtained bounds of
$O(n^2/\log^2 n)$ up to factors polynomial in $\log\log n$ for
both distances.
In one dimension, Blank and Driemel~\cite{BD24} obtained an exact
algorithm for the Fr\'echet distance in
$O(n\log n+m^2\log^2 n)$ time. Their result takes the form of a
distance oracle: after $O(n\log n)$ preprocessing of the $n$-vertex curve,
queries with an $m$-vertex curve take $O(m^2\log^2 n)$ time.
A query may also specify a subcurve of the preprocessed curve.

For conditional lower bounds, Bringmann~\cite{Bri14} showed that
SETH rules out strongly subquadratic algorithms for both distances in
the plane, even with a sufficiently small constant approximation
factor and for imbalanced inputs. Bringmann and Mulzer~\cite{BM16}
extended this hardness to the discrete Fr\'echet distance in one
dimension. Buchin, Ophelders, and Speckmann~\cite{BOS19} ruled out
strongly subquadratic approximation within any fixed factor below three
for both distances on one-dimensional curves of comparable sizes. Blank~\cite{Blank26}
strengthened the approximation lower bounds for imbalanced inputs,
with thresholds depending on the dimension and metric.
For geometrically restricted inputs, Driemel, Har-Peled, and
Wenk~\cite{DHW12} introduced $c$-packed curves and gave a
$(1+\eta)$-approximation in $\tO(c(m+n)/\eta)$ time. Bringmann and
K\"unnemann~\cite{BK17} improved this to
$\tO(c(m+n)/\sqrt{\eta})$, matching a conditional lower bound in
sufficiently high fixed dimension up to lower-order factors. Blank,
Driemel, Har-Peled, and Richter~\cite{BDHR26} obtained linear time
for fixed $c$ and $\eta$, for both the Fr\'echet distance and the
discrete Fr\'echet distance.
For arbitrary curves, Bringmann and Mulzer~\cite{BM16} gave an
$O(\alpha)$-approximation of the discrete Fr\'echet distance on two
$n$-vertex curves in $O(n\log n+n^2/\alpha)$ time, for
$1\le\alpha\le n$. Chan and Rahmati~\cite{CR18} improved the time
to $O(n\log n+n^2/\alpha^2)$ for
$1\le\alpha\le\sqrt{n/\log n}$. In particular, this gives
$O(n\log n)$ time with approximation factor $O(\sqrt{n/\log n})$.
For the Fr\'echet distance, Colombe and Fox~\cite{CF21} obtained an
$O(\alpha)$-approximation in $O((n^3/\alpha^2)\log n)$ time for
$\sqrt n\le\alpha\le n$, giving an $O(n)$-approximation in
$O(n\log n)$ time.
Van der Horst, van Kreveld, Ophelders, and Speckmann~\cite{HKS23}
extended the $O(\alpha)$-approximation to all $1\le\alpha\le n$,
with running time $O((n+nm/\alpha)\log^3 n)$. For two $n$-vertex
curves, van der Horst and Ophelders~\cite{HO24} obtained
$O((n^2/\alpha)\log^2 n)$ time using a curve simplification called
truncated smoothing. The combined journal version~\cite{HKS25}
gives $O((n^2/\alpha)\log n)$ time for two $n$-vertex curves,
with a stronger trade-off in one dimension.

Cheng, Huang, and Zhang~\cite{CHZ} gave the first constant-factor
approximation in strongly subquadratic time for arbitrary curves.
Their randomized $(7+\eta)$-approximation takes
$O(nm^{0.99}\log n)$ time for both distances and succeeds with
high probability. Liu and Wang~\cite{LW26} improved the factor
to $5+\eta$. Their randomized algorithms take $\tO(nm^{8/9})$ time
for the Fr\'echet distance and $\tO(nm^{4/5})$ time for the discrete
Fr\'echet distance.
Blank~\cite{Blank26} obtained a $(3+\eta)$-approximation for both
distances in $O((n+m^2)\log n)$ time, which is near-linear when
$m\le\sqrt n$. The same work gives a $2$-approximation of the
discrete Fr\'echet distance in one dimension in
$O(n\log n+m^2\log m)$ time. Van der Horst and
Ophelders~\cite{HO} obtained a deterministic $(3+\eta)$-approximation
in $\tO(nm^{2/3})$ time for both distances. Their framework also
covers paths in more general metric spaces under suitable free-space
assumptions, and uses preprocessing to answer reachability queries.

For edit distance, Chakraborty, Das, Goldenberg, Kouck\'y, and
Saks~\cite{CDGKS18} gave the first constant-factor approximation in
strongly subquadratic time, namely $\tO(n^{12/7})$ for strings of
length $n$. Andoni and Shekel Nosatzki~\cite{AN} improved this to
$O(n^{1+\eps})$ time for every fixed $\eps>0$, with a constant
factor depending on $\eps$. Their use of successively improved
distance oracles for substrings inspired our approach.